\pdfoutput=1
\documentclass[10pt, journal, twocolumn]{IEEEtran}
\IEEEoverridecommandlockouts
\usepackage{cite}
\usepackage[tbtags]{amsmath}
\usepackage{amssymb,amsfonts}
\usepackage{algorithmic}
\usepackage{graphicx, epstopdf}
\usepackage{textcomp}
\usepackage{xcolor}

\def\BibTeX{{\rm B\kern-.05em{\sc i\kern-.025em b}\kern-.08em
    T\kern-.1667em\lower.7ex\hbox{E}\kern-.125emX}}
\usepackage{tabularx}
\usepackage{bm}
\usepackage{svg, pdfpages}
\usepackage{caption}
\usepackage{subcaption}
\usepackage{lettrine}
\usepackage{comment}
\usepackage{soul}
\usepackage{makecell}

\title{Channel Estimation in RIS-Enabled mmWave Wireless Systems: A Variational Inference Approach}

\newcommand{\white}{\textcolor{white}{.}}
\newcommand{\bb}[1]{\mathbb{#1}}
\newcommand{\diag}{\text{diag}}
\newcommand{\trace}{\text{Tr}}
\newcommand{\BS}{\text{BS}}
\newcommand{\RIS}{\text{RIS}}

\author{Firas Fredj, Amal Feriani, Amine Mezghani, and Ekram Hossain\thanks{F. Fredj, A. Feriani, A. Mezghani, and E. Hossain are with the Department of Electrical and Computer Engineering at the University of Manitoba, Canada (email: \{fredjf1, feriania\}@myumanitoba.ca, \{amine.mezghani, ekram.hossain\}@umanitoba.ca).
A part of the work was presented in IEEE ICC’23 \cite{conference}.
}}
\begin{document}

\maketitle

\begin{abstract}
Channel estimation in reconfigurable intelligent surfaces (RIS)-aided systems is crucial for optimal configuration of the RIS and various downstream tasks such as user localization. In RIS-aided systems, channel estimation involves estimating two channels for the user-RIS (UE-RIS) and RIS-base station (RIS-BS) links. In the literature, two approaches are proposed: (i) cascaded channel estimation where the two channels are collapsed into a single one and estimated using training signals at the BS, and (ii) separate channel estimation that estimates each channel separately either in a passive or semi-passive RIS setting. In this work, we study the separate channel estimation problem in a fully passive RIS-aided millimeter-wave (mmWave) single-user single-input multiple-output (SIMO) communication system. First, 
we adopt a variational-inference (VI) approach to jointly estimate the UE-RIS and RIS-BS instantaneous channel state information (I-CSI). In particular, auxiliary posterior distributions of the I-CSI are learned through the maximization of the evidence lower bound. However, estimating the I-CSI for both links in every coherence block results in a high signaling overhead to control the RIS in scenarios with highly mobile users. 
Thus, we extend our first approach to estimate the slow-varying statistical CSI of the UE-RIS link overcoming the highly variant I-CSI. Precisely, our second method estimates the I-CSI of RIS-BS channel and the UE-RIS channel covariance matrix (CCM) directly from the uplink training signals in a fully passive RIS-aided system.
The simulation results demonstrate that using maximum a posteriori channel estimation using the auxiliary posteriors can provide a capacity that approaches the capacity with perfect CSI. Leveraging the UE-RIS CCM enhances spectral efficiency by minimizing the training overhead required to control the RIS, and exploiting its low-rank structure reduces training overhead compared to the maximum likelihood estimator.
\end{abstract}

\section{Introduction}
\lettrine{M}{illimeter-wave} (mmWave) communication is one of the emerging technologies for 5G/6G communication systems and beyond to meet the high data rate and spectral efficiency requirements \cite{saad2019vision6G}. Although mmWave communications offer a significant gain in throughput thanks to the increased available bandwidth, they are more susceptible to blockages due to rapid signal attenuation and severe path loss. 
In this context, reconfigurable intelligent surfaces (RISs) have been proposed to mitigate the challenges in mmWave communication systems and also enable smart and reconfigurable wireless environments \cite{zheng2022survey, dang20206G}. A RIS is a two-dimensional (2D) array consisting of a large number of passive low-cost reflecting elements that redirect the impinging electromagnetic waves following a specific phase shift pattern to create a favorable environment for the propagation of the signals\cite{nadeem2019intelligent, shao2022target}. By manipulating the signals' phases and amplitudes, the RIS can create constructive or destructive interference, amplify or attenuate the signals, and improve the communication link quality and coverage \cite{pei2021ris}.
This technology has many potential benefits, including improving the signal-to-noise ratio (SNR), increasing coverage and capacity, reducing power consumption, and enhancing security and privacy \cite{liu2021ris_survey,you2020energy,staat2021intelligent}.
In contrast to non-regenerative relays (also called repeaters), the RIS operates efficiently in full-duplex without self-interference or noise amplification \cite{di2020reconfigurable,nandan2022intelligent}. As a passive structure, the RIS introduces no additional noise beyond the environmental thermal noise level, similar to other passive scattering objects in the system. This stands as a notable advantage over active repeaters \cite{guo2022comparison}.

To achieve the desired performance through passive and active beamforming, it is crucial to accurately estimate the channel state information (CSI) between the RIS and the transceivers  \cite{wu2019reflectbeamforming, huang2019energyefficient}. 
This is a challenging problem since (i) passive RISs are unable to transmit or receive training sequences, restricting the estimation to the pilot signals at the receiver, and (ii) the number of channel coefficients to estimate increases with the number of RIS elements, limiting the feasibility of CSI acquisition within a practical channel coherence time.

\subsection{Related Work}

\noindent\textbf{Cascaded channel estimation} focuses on estimating the channel between the user equipment (UE) and the base station (BS) through the RIS (UE-RIS-BS) from the training signal received at the BS.
For instance, a compressed sensing-based method, exploiting the sparse structure of the channels, was proposed for a single-user narrowband setup \cite{wang2020compressed}. Additionally, a channel estimation scheme was developed for an RIS-aided multi-user broadband communication system by leveraging the shared channel between the RIS and BS (RIS-BS) among the users, which improves the training efficiency \cite{zheng2020intelligent}. 
In mmWave communication, the channel has a low-rank structure and is modeled by using a small number of paths compared to the number of antennas at the transceivers. Each path has a direction of departure (DoD) and a direction of arrival (DoA). For the high dimensional RIS-BS and UE-RIS channels, a two-stage non-iterative downlink channel estimation framework can be adopted by first estimating the DoDs and DoAs for the RIS-BS and UE-RIS links, respectively, and then calculating the cascaded channel using the obtained DoDs and DoAs from the first stage \cite{ardah2021trice}.
Several data-driven techniques have been also proposed to solve the cascaded channel estimation problem \cite{liu2021deep, shen2023deep, liu2020deep, zhang2021deep}. 
For instance, a deep residual learning-based approach was adopted to denoise the least square (LS) estimates by exploiting their spatial features with a convolutional neural network (CNN) \cite{liu2021deep}. However, the LS estimator suffers from high training overhead due to the large number of channel coefficients to estimate.
Addressing this shortcoming, previous work combines the super-resolution CNNs with deep denoising CNNs (DnCNNs) to estimate the cascaded channel and denoise the estimates in a MIMO OFDM communication system \cite{shen2023deep}.
For semi-passive RIS, wherein a limited number of active elements are deployed, a hybrid method used compressed-sensing to estimate the cascaded channel coefficients from a low-resolution channel matrix and a DnCNN to further denoise and improve the estimation quality \cite{liu2020deep}.
Another line of work trained a neural network to compute the optimal locations of the active RIS elements, afterward the full channel matrix was extrapolated from the estimated channels of the selected active antennas using a CNN \cite{zhang2021deep}.

The knowledge of the cascaded channel enables the RIS configuration and optimal precoding. However, this approach has various drawbacks: (i) it is not suitable for user tracking due to the coupling of DoDs and DoAs at the RIS \cite{zegrar2020userTracking,palmucci2023userTrackingNF}, and (ii) it does not exploit the slow-varying feature of the RIS-BS channel to reduce the training overhead \cite{zheng2022survey}. Acquiring the RIS-BS and UE-RIS channels separately overcomes these limitations as it decouples the cascaded channel and enables the identification of the channels' characteristics in each link.

\noindent\textbf{Separate channel estimation} has gained attention in the existing literature.
The decomposition of the cascaded UE-RIS-BS channel into two separate channels has been studied in RIS-aided systems with fully passive RIS elements. 
It was shown that the received signal follows the parallel factor tensor model which is used to develop an iterative alternating estimation scheme to obtain estimates of the UE-RIS and RIS-BS channels separately based on the Khatri-Rao factorization of the cascaded channel \cite{de2021parafac}.
However, the training overhead is still considerably high for a fully passive RIS. The use of semi-passive setup with active sensing elements at the RIS was proposed to estimate the RIS-BS channels as an initial step. Then, using the slow-varying property of the RIS-BS channel, only the UE-RIS channel is estimated in the training time of the subsequent coherence blocks \cite{hu2021semi}.
In the same context of semi-passive RISs, a variational inference (VI)-based method was developed to reduce the training overhead and estimate the channels using only the uplink training signals\cite{kim2022bayesian}. Different from this work, we propose a VI-based approach for separate channel estimation for \emph{fully passive} RIS.

Again, the aforementioned works focused on estimating the I-CSI of either the cascaded channel or the separate channels.
Estimating the I-CSIs is practical for static users' scenarios, however, it can be impractical in scenarios with high user mobility and large RISs.
Although the use of the instantaneous channels may lead to optimal phase shift configuration, it is a challenging task in practice. First, the coherence time of the mmWave channels can be drastically shorter than that in sub-6GHz channels \cite{akdeniz2014millimeter}, in particular for high mobile users. Hence, the channel estimation and phase optimization need to be performed repeatedly after every coherence time of the UE-RIS channel link, which will entail a significant amount of training overhead and tremendous computational resources accompanied by spectral inefficiency due to the pilots sent in each coherence block. Furthermore, the system optimization based on the I-CSI requires frequent transmissions of control signals from the BS to the RIS, which involves a considerable amount of signaling overhead.

\noindent\textbf{Statistical CSI} (S-CSI) has recently emerged as an essential approach in addressing the active and passive beamforming in RIS-assisted wireless systems reducing the overhead of the channel estimation and extending the coverage for practical use \cite{he2022ris,xu2023reconfiguring}.
For example, S-CSI was employed in a two-timescale beamforming design to reduce the training overhead and signal processing for acquiring the I-CSI with a specific transmission protocol \cite{han2019large}. The main idea relies on optimizing the phase-shifts based on the S-CSI  while computing the downlink beamforming vectors based on the I-CSI of the effective channel between the UEs and the BS through the RIS (i.e., UE-RIS-BS channel including the phase-shifts optimized).
A more sophisticated algorithm was proposed in \cite{zhao2020tts} to cover a more general fading channel with discrete phase-shifts in both single-user and multi-user cases. In mmWave scenarios, the S-CSI was exploited for joint hybrid and passive precoder design using block-coordinate descent-based algorithms to maximize the ergodic capacity \cite{yang2020stats_mmwave}.
However, for the RIS-aided systems, the problem of direct S-CSI estimation has not been well studied in the literature. Typically, the S-CSI is characterized by the spatial channel covariance matrix (CCM) \cite{park2016spatial}. The estimation of the spatial CCM is challenging since the complexity increases as a function of the number of RIS elements.
To address this problem, a CCM estimation method for the cascaded UE-RIS-BS channel was proposed in \cite{wang2023covariance_estimation} by exploiting the low-rank and the semi-definite three-level Toeplitz structure of the covariance matrix.
Table \ref{tab:related_work} summarizes several works in the area of I-CSI and S-CSI estimation in RIS-aided systems.

\begin{table*}[ht!]\scriptsize
\centering
\caption{Summary of channel estimation methods in RIS-aided systems}
\label{tab:related_work}
    \begin{tabular}{|l|l|l|l|}
        \hline
        Ref & Main contribution & Type of RIS & Type of estimated CSI \\
        \hline \hline
        \cite{wang2020compressed,ardah2021trice} & Cascaded channel estimation based exploiting low-rank structure of mmWave channels & Passive & Cascaded I-CSI \\ \hline
        \cite{zheng2020intelligent} & Cascaded channel estimation for multi-user setting in OFDMA system  & Passive & Cascaded I-CSI \\ \hline
        \cite{liu2021deep, shen2023deep} & Cascaded channel estimation using hybrid supervised DL techniques to denoise estimates  & Passive & Cascaded I-CSI \\ \hline
        \cite{liu2020deep, zhang2021deep} & Cascaded channel estimation based on supervised CNNs to improves estimates & Semi-passive & Cascaded I-CSI \\ \hline
        \cite{de2021parafac} & Separate channel estimation based on factorization/decomposition of the cascaded channel & Passive & Cascaded I-CSI \\ \hline
        \shortstack{\cite{hu2021semi} \\ \white} & \shortstack[l]{\white \\ Separate channel estimation based on signal parameters via rotation invariance technique \\ (ESPRIT) and multiple signal classification (MUSIC)} & \shortstack{Semi-passive \\ \white} & \shortstack{Separate I-CSI \\ \white} \\ \hline
        \cite{kim2022bayesian} & Separate channel estimation using VI-sparse Bayesian learning relying on uplink training signal & Semi-passive & Separate I-CSI\\ \hline
        \shortstack{\cite{wang2023covariance_estimation} \\ \white} &\shortstack[l]{\white \\ Cascaded channel covariance estimation based exploiting low-rank and 3-level Toeplitz structure \\ of the covariance matrix} & \shortstack{Passive RIS \\ \white} & \shortstack{Cascaded S-CSI \\ \white} \\ \hline
        \shortstack{\textbf{This paper} \\ \white \\ \white} & \shortstack[l]{\white \\ This paper proposes amortized VI to separately estimate in mmWave communication \\(i) I-CSI of UE-RIS and RIS-BS channels,\\ (ii) I-CSI of RIS-BS channel and S-CSI of UE-RIS channel} & \shortstack{Passive \\ \white \\ \white} & \shortstack[l]{(i) Separate I-CSI \\ (ii) Hybrid: I-CSI \\ and S-CSI } \\ \hline
    \end{tabular}
\end{table*}

\subsection{Motivation and Contributions}
To overcome the challenges discussed in the previous section, our work focuses on separate channel estimation in a fully passive RIS-aided network. Again, the separate channel estimation incurs less training overhead channel estimation schemes, as the RIS-BS channel is slow-varying, compared to cascaded channel estimation and the fully-passive RIS setup has lower power consumption compared to the semi-passive one. Consequently, our solution relies only on the uplink training signal at the BS to estimate both channel matrices.

From a Bayesian inference perspective, the acquisition of the posterior distribution of the separate channels becomes challenging due to the passive nature of the RIS.
Therefore, we advocate the utilization of a VI-based framework providing an approximation of the intractable posterior distribution with convenient distributions. Diverging from conventional deterministic models, VI introduces a probabilistic paradigm that seamlessly integrates uncertainties allowing the incorporation of prior information.
First, we propose a joint channel estimation (JCE) method where the intractable posterior distribution of the UE-RIS and RIS-BS channels are approximated by complex Laplace auxiliary distributions. We employ the \emph{amortized VI} framework where neural networks are used to map the training signals to the parameters of the auxiliary distributions. These neural networks are trained to maximize the \emph{evidence lower bound} (ELBO), an equivalent objective to minimizing the Kullback-Leibler (KL) divergence between the true posterior distribution of the channels and the auxiliary distributions. Then, using the predicted parameters, we employ the maximum a posteriori (MAP) to estimate the channels.

Optimizing the phase-shifts according to the I-CSI can incur substantial signaling overhead at the RIS. This arises from the necessity to update the RIS configuration in each coherence block, particularly inconvenient when considering the rapid and dynamic changes of the UE-RIS channel.
To reduce the signaling overhead, we propose to estimate the UE-RIS CCM instead of the UE-RIS I-CSI. We keep estimating the RIS-BS I-CSI since the RIS-BS channel varies slowly compared to the dynamic UE-RIS channel.
This leads to our second approach, coined joint channel-covariance estimation (JCCE), that extends the use of the VI-based framework to directly estimate the RIS-BS I-CSI and UE-RIS CCM from the received training signal at BS. 
In particular, JCCE implements the VI-based framework to effectively approximate the posterior distributions of the RIS-BS channel and the UE-RIS CCM. Similar to the methodology applied in the JCE method, we leverage the auxiliary distributions, whose parameters are predicted by the neural networks, to obtain the MAP estimates. Considering the large size of the UE-RIS CCM resulting from the large number of elements at the RIS, we exploit the inherent low-rank structure of the covariance matrix of the mmWave channels.
Different from the traditional methods, our approach directly estimates the UE-RIS CCM from the training signals, eliminating the need for multiple intermediary channel estimation steps before the CCM computation. Also,
unlike previous art where the covariance matrix of the cascaded channel was estimated \cite{wang2023covariance_estimation}, our approach estimates the RIS-BS channel and UE-RIS CCM separately. Finally, we derive the phase-shifts in closed form, that aims at maximizing the capacity based on the RIS-BS channel and the UE-RIS CCM.

Our solutions are flexible and take into account the sparsity of mmWave channels as they do not require foreknowledge of the number of paths prior to the estimation process. {\em The proposed solutions can also be extended to other types of channels}.
To summarize, our major contributions are as follows:
\begin{enumerate}
    \item Using VI-based framework, we separately estimate the I-CSI in an RIS-aided mmWave systems with fully-passive RIS elements by learning the auxiliary distributions that approximate the true posteriors of the channels;
    \item We extend our first approach to estimate the slow-varying RIS-BS channel and the UE-RIS CCM, which are used for RIS phase-shift optimization over several coherence blocks;
    \item We develop a closed-form expression for the phase-shifts that optimize the transmission capacity given the estimates of the RIS-BS channel and the UE-RIS CCM;
    \item We demonstrate the effectiveness of our proposed methods by comparing the achieved capacity with the capacity obtained using the perfect CSI. An improvement in spectral efficiency is shown while using the JCCE method compared to the JCE in addition to the substantial signal complexity reduction inherited by relying on the slow-varying RIS-BS channel and UE-RIS CCM for the passive beamforming.
\end{enumerate}

\subsection{Paper Organization and Notation}
The remainder of this paper is organized as follows: Section \ref{sec:system} describes the system model, and details the VI-based framework for estimation as well as the variational neural networks. Section III presents the proposed VI-based methods for joint RIS-BS and UE-RIS channel estimation, and the joint RIS-BS channel and UE-RIS CCM estimation. Section IV presents the derivations of the closed-form expressions for RIS  phase-shifts based on the channel estimates. In Section V, we describe the simulation setup, and we present the numerical results in Section VI before we conclude the paper.

The list of symbols that will be later used in the paper is given in Table~\ref{tab:list_symbols}. Scalars, vectors and matrices are denoted by $x$, $\bm{x}$, and $\bm{X}$, respectively. $\bm{X}^*$ and $\bm{X}^{\mathsf{H}}$ denote the complex conjugate and conjugate transpose of $\bm{X}$. The $i$-th element of a vector $\bm{x}$ is $\bm{x}_i$, while the $(i,j)$-th element of a matrix $\bm{X}$ is $\bm{X}_{i,j}$. The $n\times n$ identity matrix is written as $\bm{I}_n$. The $\diag(\bm{x})$ is the diagonal matrix with the elements of the vector $\bm{a}$ on the main diagonal. The element-wise product of $\bm{X}$ and $\bm{Y}$ is written as $\bm{X} \circ \bm{Y}$, while the Khatri-Rao product between $\bm{X}$ and $\bm{Y}$ is written as $\bm{X}\odot \bm{Y}$. $\bm{X} \otimes \bm{Y}$ denotes the kronecker product between $\bm{X}$ and $\bm{Y}$. $\trace(\bm{X})$ and $|\bm{X}|$ represent the trace and determinant of the matrix $\bm{X}$, respectively, and $|x|$ represents the absolute value of a complex number $x$, while $\angle x$ is the phase of $x$.  The complex Gaussian random vector is denoted as $\bm{x}\sim \mathcal{CN}(\bm{m},\bm{\Sigma})$ with mean $\bm{m}$ and covariance matrix $\bm{\Sigma}$, whereas a complex Laplace random variable $x$ is denoted as $x\sim \mathcal{CL}(m,b)$ with mean $m$, scale $b$ and probability density function (PDF) given by:
\begin{equation}
    p(x) = \frac{1}{2\pi b^2} e^{-\frac{|x-m|}{b}}.
\end{equation}
A Gamma distributed random variable with unit scale is denoted as $x\sim \text{Gamma}(k)$ with shape $k$, while an Exponentially distributed random variable with rate $\alpha$ is denoted by $x \sim \text{Exp}(\alpha)$.

\begin{table}[!t]\scriptsize
    \centering
    \caption{List of symbols}
    \label{tab:list_symbols}
    \begin{tabularx}{0.5\textwidth}{lX}
    \hline
    \multicolumn{2}{c}{\textbf{System model}}\\
    \hline
    $\rho$ & Signal-to-noise ratio (SNR) \\
    $M, N$ & Number of BS antennas and RIS elements \\
    $N_p$ & Number of pilots per UE-RIS coherence block\\
    $N_b$ & Number of coherence blocks used for training\\
    $Q, P$ & Number of paths of UE-RIS and RIS-BS channels \\
    $\bm{v}$ & The phase-shifts vector\\
    $\bm{h},\bm{G}$ & UE-RIS and RIS-BS channels in the time domain\\
    $\bm{h}^\mathsf{vir}, \bm{G}^\mathsf{vir}$ & UE-RIS and RIS-BS channels in the angular domain\\
    $\bm{R}_{\bm{h}}, \bm{d}$ & The covariance matrix and angular correlation vector of the UE-RIS link\\
    $\bm{\Phi}$ & RIS configuration used for uplink training\\
    $\bm{F}_{N}, \bm{F}_{M}$ & Discrete Fourier Transform matrices (DFT)\\
    \hline
    \multicolumn{2}{c}{\textbf{Variational Inference}}\\
    \hline
    $\mathcal{L}^\mathsf{I-CSI},\mathcal{L}^\mathsf{S-CSI}$ & ELBO functions for the I-CSI and S-CSI cases\\
    $p(\bm{h},\bm{G}|\bm{Y})$ & True posterior of the UE-RIS and RIS-BS channels\\
    $q_{\bm{\lambda}_1}(\bm{h}^\mathsf{vir}|\bm{Y})$ & Auxiliary posterior of UE-RIS channel in the angular domain with statistical parameters $\bm{\lambda}_1$\\
    $q_{\bm{\lambda}_2}(\bm{G}^\mathsf{vir}|\bm{Y})$ & Auxiliary posterior of RIS-BS channel in the angular domain with statistical parameters $\bm{\lambda}_2$\\
    $q_{\bm{\lambda}_1}(\bm{d}|\bm{Y})$ & Auxiliary posterior of the angular correlation vector with statistical parameters $\bm{\lambda}_1$\\
    $p(\bm{h}^\mathsf{vir})$ & Prior of the UE-RIS channel in the angular domain\\
    $p(\bm{G}^\mathsf{vir})$ & Prior of the RIS-BS channel in the angular domain\\
    $p(\bm{d})$ & Prior of the angular correlation\\
    $\mathcal{F}$, $\mathcal{G}$ & The encoder networks trained to predict the statistical parameters $\bm{\lambda}_1$ and $\bm{\lambda}_2$, respectively\\
    $\bm{\mathcal{W}}_1$, $\bm{\mathcal{W}}_2$ & Weights of encoders $\mathcal{F}$ and $\mathcal{G}$, respectively \\
    \hline
    \end{tabularx}
\end{table}

\section{System Model, Assumptions, and Variational Inference Approach}\label{sec:system}

\subsection{System Model, Assumptions, and Methodology}
\begin{figure}[t!]
    \centering
    \includegraphics[width=\linewidth]{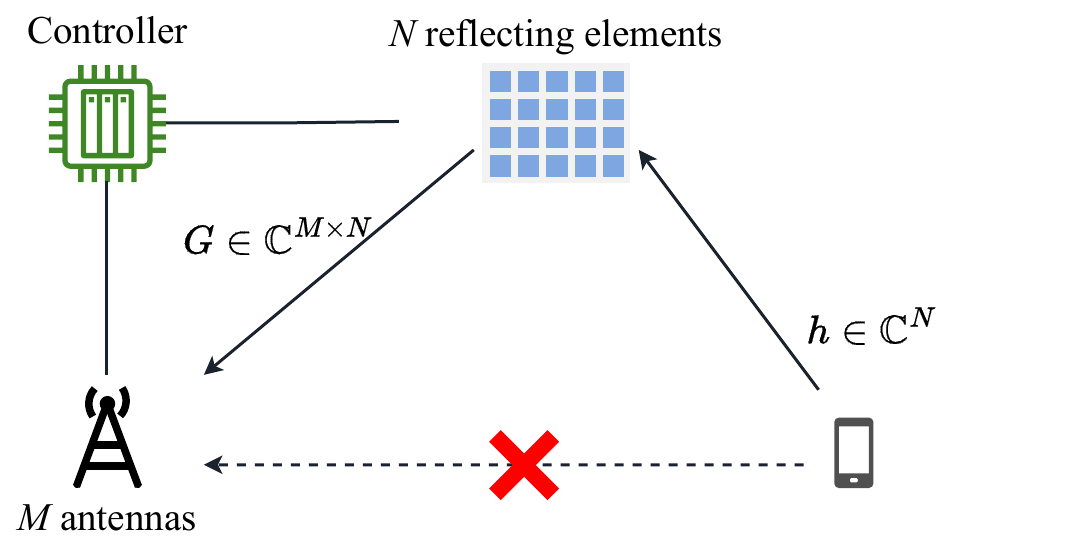}
    \caption{RIS-aided wireless communication system.}
    \label{fig:irs}
\end{figure}

We consider a RIS-assisted single-user communication system with $M$ antennas at the BS, $N$ passive reflecting elements at the RIS, and a single antenna at the user, as illustrated in Fig. \ref{fig:irs}.
Considering the uplink transmission, the UE-RIS and RIS-BS channels are denoted by $\bm{h}\in \mathbb{C}^{N}$ and $\bm{G} \in \mathbb{C}^{M \times N}$, respectively. 
We assume the direct link between the UE and the BS is blocked.
Furthermore, we adopt a block-fading channel model where the RIS-related channels $\bm{G}$ and $\bm{h}$ are considered quasi-static within a coherence time denoted by $T_{\bm{G}}$ and $T_{\bm{h}}$, respectively.
Hence, the received signal at the BS can be expressed as follows:
\begin{equation}
    \bm{y} = \sqrt{\rho}\; \bm{G}\; \text{diag}(\bm{v})\; \bm{h}\;x + \bm{w},
\end{equation}
where $\rho$, $x \in \mathbb{C}$, and $\bm{w}\in\mathbb{C}^M$ are the SNR, the transmitted signal, and the additive white noise (i.e., $\bm{w} \sim \mathcal{CN}(\bm{0}, \bm{I}_M)$), respectively. The phase shifts contributed by the RIS are represented by the diagonal matrix $\text{diag}(\bm{v})$, where $\bm{v}=[e^{j\theta_1},\dots,e^{j\theta_N}]^T$ with $\theta_n \in [0,2\pi)$ being the phase shift of the $n$-th element in the RIS.

In the next section, we will describe the VI-based approach which will be applied in the subsequent sections to solve the joint RIS-BS and UE-RIS channel estimation, and the joint RIS-BS channel and UE-RIS CCM estimation problems.

\subsection{Variational Inference (VI) Approach}

The variational methods are a class of systematic approaches that approximate complex and intractable probability distributions with convenient tractable ones. VI is a specific case of variational methods that infers the marginal distributions or likelihood functions of hidden variables in a statistical model \cite{tzikas2008variational} \cite{blei2017variational}. For instance, we consider a communication model with two unknown inputs denoted $\bm{z}_1$ and $\bm{z}_2$ (e.g., RIS-BS and UE-RIS channels) and an observed output $\bm{Y}$, and we assume that the output is obtained following a certain probability $p(\bm{Y}|\bm{z}_1,\bm{z}_2)$. If the goal is to infer $\{\bm{z}_1,\bm{z}_2\}$ based on the evidence $\bm{Y}$, we have interest in deriving the probability $p(\bm{z}_1,\bm{z}_2|\bm{Y})$. When the direct evaluation of the posterior distribution $p(\bm{z}_1,\bm{z}_2|\bm{Y})$ is infeasible, VI allows us to approximate the posterior $p(\bm{z}_1,\bm{z}_2|\bm{Y})$ with a parameterized tractable distribution $q_{\bm{\lambda}}(\bm{z}_1,\bm{z}_2|\bm{Y})$.

The central concept in VI is the Evidence Lower Bound (ELBO), also known as the variational lower bound. It serves as a surrogate for the intractable log-likelihood of the data, and maximizing it corresponds to minimizing the Kullback-Leibler (KL) divergence between the true posterior $p(\bm{z}_1,\bm{z}_2|\bm{Y})$ and the variational approximation $q_{\bm{\lambda}}(\bm{z}_1,\bm{z}_2|\bm{Y})$.
The ELBO is given by \cite{zhang2018advances}:
\begin{equation}
\begin{split}
    \log p(\bm{Y}) &\geq \mathbb{E}_{\bm{z}_1,\bm{z}_2 \sim q_{\bm{\lambda}}(\bm{z}_1,\bm{z}_2|\bm{Y})} \left[\log \frac{p(\bm{z}_1,\bm{z}_2,\bm{Y})} {q_{\bm{\lambda}}(\bm{z}_1,\bm{z}_2|\bm{Y})} \right] \\
    &\triangleq -\mathcal{L}(\bm{Y};\bm{\lambda}).
\end{split}
\end{equation}

Assuming that $q_{\bm{\lambda}}(\bm{z}_1,\bm{z}_2|\bm{Y})$ belongs to a family of tractable distributions, the VI approach optimizes the parameters $\bm{\lambda}$ of the approximated distribution $q_{\bm{\lambda}}(\bm{z}_1,\bm{z}_2|\bm{Y})$ such that the objective function $\mathcal{L}(\bm{Y};\bm{\lambda})$ is minimized.

We further assume that the approximated distribution can be factorized as $q_{\bm{\lambda}}(\bm{z}_1,\bm{z}_2|\bm{Y})=q_{\bm{\lambda}_1}(\bm{z}_1|\bm{Y})\cdot q_{\bm{\lambda}_2} (\bm{z}_2|\bm{Y})$ where $\bm{\lambda}=(\bm{\lambda}_1,\bm{\lambda}_2)$ and we optimize the independent distributions by minimizing $\mathcal{L}(\bm{Y};\bm{\lambda}_1,\bm{\lambda}_2)$. 
This independence assumption is referred to as the \textit{mean-field approximation} \cite{blei2017variational}. It is equivalent to assuming a low correlation between $\bm{z}_1$ and $\bm{z}_2$ conditioned on $\bm{Y}$. Hence, the objective function is simplified to a general form given by:
\begin{equation}\label{eq:elbo_general_form}
\begin{split}
    \mathcal{L}(\bm{Y}; \bm{\lambda}_1,\bm{\lambda}_2) &=\begin{aligned}[t] &\underbrace{\mathbb{E}_{\bm{z}_1 \sim q_{\bm{\lambda}_1}(\bm{z}_1|\bm{Y})} \left[\log \frac{q_{\bm{\lambda}_1}(\bm{z}_1|\bm{Y})} {p(\bm{z}_1)} \right]}_{\mathcal{L}_1}\\
    &\!\! +\underbrace{\mathbb{E}_{\bm{z}_2 \sim q_{\bm{\lambda}_2}(\bm{z}_2|\bm{Y})} \left[\log \frac{q_{\bm{\lambda}_2}(\bm{z}_2|\bm{Y})} {p(\bm{z}_2)} \right]}_{\mathcal{L}_2}\\
    \end{aligned}\\
    &\quad \underbrace{- \mathbb{E}_{\bm{z}_1,\bm{z}_2 \sim q_{\bm{\lambda}}(\bm{z}_1,\bm{z}_2|\bm{Y})} \bigl[\log p(\bm{Y}|\bm{z}_1,\bm{z}_2)\bigl]}_{\mathcal{L}_3}.
\end{split}
\end{equation}

Note that $\mathcal{L}_1$ and $\mathcal{L}_2$ in Eq. (\ref{eq:elbo_general_form}) represent the KL divergence between the auxiliary distributions, also known as variational distributions, $q_{\bm{\lambda}_1}(\bm{z}_1|\bm{Y})$ and $q_{\bm{\lambda}_2}(\bm{z}_2|\bm{Y})$ and their actual priors $p(\bm{z}_1)$ and $p(\bm{z}_2)$, respectively. Regarding $\mathcal{L}_3$, it corresponds to the reconstruction error of the estimated pilot signal $\bm{\widehat{Y}}$ with the auxiliary distributions $q_{\bm{\lambda}_1}(\bm{z}_1|\bm{Y})$ and $q_{\bm{\lambda}_2}(\bm{z}_2|\bm{Y})$. Hence, minimizing the objective function $\mathcal{L} = \mathcal{L}_1 + \mathcal{L}_2 + \mathcal{L}_3$ ensures that the generated posterior distributions are close to the prior distributions and the reconstructed signal $\bm{\widehat{Y}}$ is similar to the received signal.

After deriving the ELBO, one common approach is to use neural networks to parameterize the approximate posterior distribution \cite{miao2016neural}. In this approach, a neural network is used to map the observed data to the parameters of the auxiliary distribution, such as the mean and the scale parameters of a complex Laplace distribution. The neural network is typically trained using stochastic gradient descent or a related optimization algorithm to minimize the KL divergence between the auxiliary distribution and the true posterior distribution, as represented by the ELBO.

\begin{figure}[!t]
    \centering
    \includegraphics[width=\linewidth]{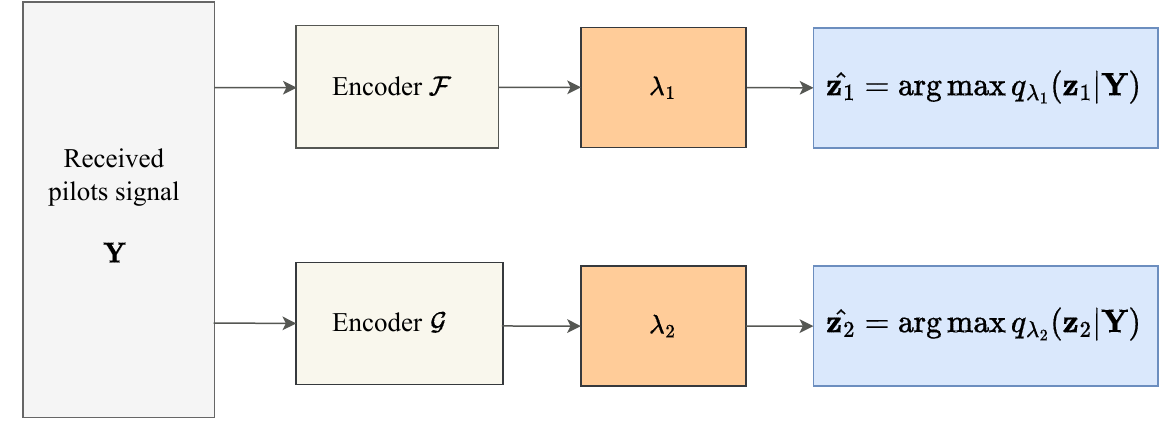}
    \caption{Variational neural networks.}
    \label{fig:neural_networks}
\end{figure}

Therefore, we obtain the parameters of the two auxiliary distributions $q_{\bm{\lambda}_2}(\bm{z}_2|\bm{Y})$ and $q_{\bm{\lambda}_1}(\bm{z}_1|\bm{Y})$ by two trainable neural networks:
\begin{align}
\label{eq:nn_eq}
    \bm{\lambda}_1 = \mathcal{F}_{\bm{\mathcal{W}}_1}(\bm{Y}); \qquad \bm{\lambda}_2 = \mathcal{G}_{\bm{\mathcal{W}}_2}(\bm{Y})
\end{align}
referred to by \textit{Encoder} $\mathcal{F}$ and \textit{Encoder} $\mathcal{G}$, as shown in Fig. \ref{fig:neural_networks}, where $\bm{\mathcal{W}}_1$ and $\bm{\mathcal{W}}_2$ are the weights of the neural networks.
In particular, the neural networks take the training signal $\bm{Y}$ which is the observed data as input and outputs the parameters of the distributions $q_{\bm{\lambda}_1}(\bm{z}_1|\bm{Y})$ and $q_{\bm{\lambda}_2}(\bm{z}_2|\bm{Y})$.
The neural networks learn to encode the data into a meaningful representation that captures the latent information. The parameters of the two neural networks \textit{Encoder} $\mathcal{F}$ and \textit{Encoder} $\mathcal{G}$ are learned by minimizing the loss function in Eq. (\ref{eq:elbo_general_form}):
\begin{equation}
    \bm{\mathcal{W}}^*_1, \bm{\mathcal{W}}^*_2 = \arg \min_{\bm{\mathcal{W}}_1,\bm{\mathcal{W}}_2} \mathcal{L}\left(\bm{Y}; \mathcal{F}_{\bm{\mathcal{W}}_1}(\bm{Y}); \mathcal{G}_{\bm{\mathcal{W}}_2}(\bm{Y})\right).
\end{equation}

\section{CSI Estimation via Variational Inference}\label{sec:JCE}
In this section, we present our proposed approaches for separate channel estimation in a mmWave wireless communication system with fully passive RIS. Our first approach, named JCE, estimates both RIS-BS and UE-RIS I-CSI using uplink training signals in an end-to-end fashion. Next, we detail our second method, called JCCE, where we estimate the UE-RIS CCM instead of the I-CSI.

\subsection{Joint Channel Estimation via Variational Inference}

We aim to estimate the RIS-BS and the UE-RIS channels, $\bm{G}$ and $\bm{h}$, based on the received training signal. The training signal is obtained by sending $N_p$ pilot signals by the user to the BS through the UE-RIS-BS channel. For different pilot transmissions, different configurations of the RIS are employed. The received training signals are given by:
\begin{equation}
    \bm{Y} = \sqrt{\rho}\; \bm{G}\; \bigl(\bm{\Phi} \circ (\bm{h x^T})\bigl) \;+\; \bm{W},
\end{equation}
where $\bm{Y}=\bigl[\bm{y}_1,\dots,\bm{y}_{N_p}\bigl] \in \mathbb{C}^{M\times N_p}$ is the concatenation of the $N_p$ training signals, $\bm{x} = \bigl[x_1,\dots,x_{N_p}]^T$ denotes the pilots sent by the user, $\bm{\Phi} = \bigl[\bm{v}_1,\dots,\bm{v}_{N_p}\bigl]$ is concatenation of the phase-shifts vectors used where $\bm{v}_l$ is assigned to the $l$-th pilot signal, and $\bm{W}=\bigl[\bm{w}_1,\dots,\bm{w}_{N_p}\bigl]$ is the noise matrix.

In mmWave communication and due to the large number of elements in the RIS and the high path loss, the channels are sparse in the angular domain \cite{zheng2022survey}. Specifically, only a small number of paths contribute to the received signal, and the other paths are negligible.
The channels in the angular domain can be obtained by applying the Discrete Fourier Transform (DFT) as follows:
\begin{align}
    \bm{G}^\mathsf{vir} = \bm{F}_M \bm{G} \bm{F}_N;\qquad\bm{h}^\mathsf{vir} = \bm{F}_N \bm{h},
\end{align}
where $\bm{F}_N$ and $\bm{F}_{M}$ are the DFT matrices of size $N\times N$ and $M\times M$, respectively. $\bm{G}^\mathsf{vir}$ and $\bm{h}^\mathsf{vir}$ are the channels in the angular domain where the elements are independent and identically distributed and distributed according to a complex Laplace distribution with zero mean and scales $\alpha_{\bm{G}^\mathsf{vir}}$ and $\alpha_{\bm{h}^\mathsf{vir}}$, respectively, i.e., $\bm{G}^\mathsf{vir}_{i,j} \sim \mathcal{CL}(0, \alpha_{\bm{G}^\mathsf{vir}})$ and $\bm{h}_{i}^\mathsf{vir} \sim \mathcal{CL}(0, \alpha_{\bm{h}^\mathsf{vir}})$. Given that $\bm{F}_N^{-1}=\frac{1}{N} \bm{F}_N^{\mathsf{H}}$ for any DFT matrix of size $N\times N$, the received training signal for the $l$-th pilot signal is expressed as follows:
\begin{equation}
\label{eq:signal_model_laplace}
    \bm{y}_l = \frac{\sqrt{\rho}}{MN^2} \bm{F}_M^{\mathsf{H}} \bm{G}^\mathsf{vir} \bm{F}_N^{\mathsf{H}} \text{diag}(\bm{v}_l)\bm{F}_N^{\mathsf{H}} \bm{h}^\mathsf{vir} x_l + \bm{w}_l, l=1,\dots,N_p.
\end{equation}

By applying the VI framework, we approximate the intractable true posterior distribution $p(\bm{h}^\mathsf{vir},\bm{G}^\mathsf{vir}|\bm{Y})$ by a tractable parameterized distribution denoted $q_{\bm{\lambda}}(\bm{h}^\mathsf{vir},\bm{G}^\mathsf{vir}|\bm{Y})$ that maximizes the ELBO. Assuming a low-correlation between the channels $\bm{h}^\mathsf{vir}$ and $\bm{G}^\mathsf{vir}$ conditioned on the training signal $\bm{Y}$, by using the mean-field approximation, the auxiliary distribution is factorized as $q_{\bm{\lambda}}(\bm{h}^\mathsf{vir},\bm{G}^\mathsf{vir}|\bm{Y}) = q_{\bm{\lambda}_1}(\bm{h}^\mathsf{vir}|\bm{Y}) \cdot q_{\bm{\lambda}_2}(\bm{G}^\mathsf{vir}|\bm{Y})$.

We assume that the auxiliary distributions follow complex Laplace distributions with independent elements:
\begin{alignat}{2}
    &q_{\bm{\lambda}_1}\bigl(\bm{h}^\mathsf{vir}_{i}|\bm{Y}\bigl) \sim \mathcal{CL}\bigl(\bm{m}_{i,j}, \bm{b}_{i}\bigl) && \forall i;\\
    &q_{\bm{\lambda}_2}\bigl(\bm{G}^\mathsf{vir}_{i,j}|\bm{Y}\bigl) \sim \mathcal{CL}\bigl(\bm{M}_{i,j}, \bm{B}_{i,j}\bigl)\qquad  && \forall i,j,
\end{alignat}
where $\bm{\lambda}_1=\{\bm{m}$, $\bm{b}\}$ and $\bm{\lambda}_2=\{\bm{M},\bm{B}\}$ are the parameters of the auxiliary distributions. Their optimal values are computed by minimizing the following loss function:
\begin{equation}\label{eq:elbo_jce}
\begin{split}
    \mathcal{L}^\mathsf{I-CSI}(\bm{\lambda}_1, \bm{\lambda}_2)=& \underbrace{\mathbb{E}_{\bm{h}^\mathsf{vir} \sim q_{\bm{\lambda}_1}(\bm{h}^\mathsf{vir}|\bm{Y})} \left[\log \frac{q_{\bm{\lambda}_1}(\bm{h}^\mathsf{vir}|\bm{Y})} {p(\bm{h}^\mathsf{vir})} \right]}_{\mathcal{L}^\mathsf{I-CSI}_1}\\
    &+ \underbrace{\mathbb{E}_{\bm{G}^\mathsf{vir} \sim q_{\bm{\lambda}_2}(\bm{G}^\mathsf{vir}|\bm{Y})} \left[\log \frac{q_{\bm{\lambda}_2}(\bm{G}^\mathsf{vir}|\bm{Y})} {p(\bm{G}^\mathsf{vir})} \right]}_{\mathcal{L}^\mathsf{I-CSI}_2} \\
    &\underbrace{- \mathbb{E}_{\bm{h}^\mathsf{vir},\bm{G}^\mathsf{vir} \sim q_{\bm{\lambda}}(\bm{h}^\mathsf{vir},\bm{G}^\mathsf{vir}|\bm{Y})} \bigl[\log p(\bm{Y}|\bm{h}^\mathsf{vir},\bm{G}^\mathsf{vir})\bigl]}_{\mathcal{L}^\mathsf{I-CSI}_3}.
\end{split}
\end{equation}
\begin{figure*}[!t]
\begin{align}\label{eq:loss_3_laplace}
    \mathcal{L}_3^\mathsf{I-CSI}({\bm{\lambda}})
    &=- \sum_{l=1}^{N_p} \mathbb{E}_{\bm{h}^\mathsf{vir},\bm{G}^\mathsf{vir} \sim q_{{\bm{\lambda}}}(\bm{h}^\mathsf{vir},\bm{G}^\mathsf{vir}|\bm{Y})} \bigl[ \log p(\bm{y}_l|\bm{h}^\mathsf{vir},\bm{G}^\mathsf{vir})\bigl]\nonumber\\
    &=\sum_{l=1}^{N_p}  \biggr[ (\bm{y}_l - \frac{\sqrt{\rho}}{MN^2}\bm{F}_M^{\mathsf{H}} \bm{M} \bm{F}_N^{\mathsf{H}} \text{diag}(\bm{v}_l) \bm{F}_N^{\mathsf{H}} \bm{m} x_l)^{\mathsf{H}} (\bm{y}_l - \frac{\sqrt{\rho}}{MN^2}\bm{F}_M^{\mathsf{H}} \bm{M} \bm{F}_N^{\mathsf{H}} \text{diag}(\bm{v}_l) \bm{F}_N^{\mathsf{H}} \bm{m} x_l)\nonumber \\
    &\quad+\frac{\rho |x_l|^2}{MN^4} \cdot \trace\bigl(\bm{\Lambda} \bm{F}_N^{\mathsf{H}} \text{diag}(\bm{v}_l) \bm{F}_N^{\mathsf{H}} \bm{Q} \bm{F}_N \text{diag}(\bm{v}_l)^{\mathsf{H}} \bm{F}_N \bigl)+\frac{\rho |x_l|^2}{MN^4} \; \trace\bigl(\bm{M}^{\mathsf{H}} \bm{M} \bm{F}_N^{\mathsf{H}} \text{diag}(\bm{v}_l) \bm{F}_N^{\mathsf{H}} \bm{Q} \bm{F}_N \text{diag}(\bm{v}_l)^{\mathsf{H}} \bm{F}_N \bigl) \nonumber\\ 
    &\quad+\frac{\rho |x_l|^2}{MN^4}\; \bm{m}^{\mathsf{H}} \bm{F}_N \text{diag}(\bm{v}_l)^{\mathsf{H}} \bm{F}_N \bm{\Lambda} \bm{F}_N^{\mathsf{H}} \text{diag}(\bm{v}_l) \bm{F}_N^{\mathsf{H}} \bm{m}\biggr]+ C_1.
\end{align}
\end{figure*}
%\end{equation}

The first loss $\mathcal{L}_1^\mathsf{I-CSI}$ is the KL-divergence between the auxiliary distribution and the prior of $\bm{h}^\mathsf{vir}$, which can be expressed as follows:
\begin{equation}
\begin{split}
    \mathcal{L}_1^\mathsf{I-CSI} ({\bm{\lambda}_1})
    &=\begin{aligned}[t]&\sum_{i=1}^N \mathbb{E}_{\bm{h}^\mathsf{vir}_i \sim q_{\bm{\lambda}_1}(\bm{h}^\mathsf{vir}_i|\bm{Y})} \bigr[\log q_{\bm{\lambda}_1}(\bm{h}^\mathsf{vir}_i|\bm{Y}) \bigr]\\
    &- \mathbb{E}_{\bm{h}^\mathsf{vir}_i \sim q_{\bm{\lambda}_1}(\bm{h}^\mathsf{vir}_i|\bm{Y})} \bigr[\log p(\bm{h}^\mathsf{vir}_i)\bigr]
    \end{aligned}\\
    &=\sum_{i=1}^N H\Bigl(q_{\bm{\lambda}_1}\bigl(\bm{h}^\mathsf{vir}_i|\bm{Y}\bigl), p\bigl(\bm{h}^\mathsf{vir}_i\bigl)\Bigl) \\
    &\quad\;\;- H\Bigl(q_{\bm{\lambda}_1}\bigl(\bm{h}^\mathsf{vir}_i|\bm{Y}\bigl)\Bigl),
\end{split}
\end{equation}
where $H\Bigl(q_{\bm{\lambda}_1}\bigl(\bm{h}^\mathsf{vir}_i|\bm{Y}\bigl)\Bigl)$ is the entropy of $q_{\bm{\lambda}_1}\bigl(\bm{h}^\mathsf{vir}_i|\bm{Y}\bigl)$ and $H^\mathsf{cross-entropy} = H\Bigl(q_{\bm{\lambda}_1}\bigl(\bm{h}^\mathsf{vir}_i|\bm{Y}\bigl), p\bigl(\bm{h}^\mathsf{vir}_i\bigl)\Bigl)$ is the cross entropy between $q_{\bm{\lambda}_1}\bigl(\bm{h}^\mathsf{vir}_i|\bm{Y}\bigl)$ and $p\bigl(\bm{h}^\mathsf{vir}_i\bigl)$. The entropy of the complex Laplace distribution is:
\begin{equation}
    H\Bigl(q_{\bm{\lambda}_1}\bigl(\bm{h}^\mathsf{vir}_i|\bm{Y}\bigl)\Bigl) = \log(2\pi \bm{b}_i^2) +2.
\end{equation}
The proof can be found in the \textbf{Appendix}. The cross-entropy between two complex Laplace distributions is given by:
\begin{align}
H^\mathsf{cross-entropy} &= 
    \log(2\pi \alpha_{\bm{h}^\mathsf{vir}}^2) + \mathbb{E}_{\bm{h}^\mathsf{vir}_i \sim q_{\bm{\lambda}_1}(\bm{h}^\mathsf{vir}_i|\bm{Y})} \left[ \frac{\big|\bm{h}^\mathsf{vir}_i \big|}{\alpha_{\bm{h}^\mathsf{vir}}}\right].
\end{align}
To compute the gradient with respect to the parameters of the auxiliary distribution of the UE-RIS channel link $q_{\bm{\lambda}_1}(\bm{h}^\mathsf{vir}_i|\bm{Y})$, we employ the reparameterization trick. This technique involves evaluating the expectation using $D$ Monte-Carlo samples where the $d$-th sample is computed by $\widehat{\bm{h}^{\mathsf{vir}}}^{(d)}=\bm{m}_i + \bm{b}_i \times CL(0,1)$ to maintain the differentiability and enabling efficient optimization through gradient-based methods.
Hence, $\mathcal{L}_1^\mathsf{I-CSI}$ is expressed as:
\begin{equation}
\begin{split}
\label{eq:L1_ICSI}
    \mathcal{L}_1^\mathsf{I-CSI}({\bm{\lambda}_1}) &= \frac{1}{D} \sum_{i=1}^N \sum_{d=1}^D \frac{\big|\widehat{\bm{h}^\mathsf{vir}_i}^{(d)}\big|}{\alpha_{\bm{h}^\mathsf{vir}}} - \sum_{i=1}^N \log(2\pi \bm{b}_i^2) \\
    &\quad+N\log(2\pi \alpha_{\bm{h}^\mathsf{vir}}^2) -2N.
\end{split}
\end{equation}
Similarly, we derive $\mathcal{L}_2^\mathsf{I-CSI}$:
\begin{align}
    \mathcal{L}_2^\mathsf{I-CSI}({\bm{\lambda}_2})
    &=\frac{1}{D} \sum_{i=1}^M \sum_{j=1}^N \sum_{d=1}^D \frac{\big|\widehat{\bm{G}^\mathsf{vir}_{i,j}}^{(d)}\big|}{\alpha_{\bm{G}^\mathsf{vir}}} - \sum_{i=1}^M \sum_{j=1}^N \log(2\pi \bm{B}_{i,j}^2)\nonumber \\
    & \quad+ NM\log(2\pi \alpha_{\bm{G}^\mathsf{vir}}^2) -2NM, \label{eq:l2_jce}
\end{align}
where the Monte-Carlo samples are computed as $\widehat{\bm{G}^\mathsf{vir}_{i,j}}^{(d)} = \bm{M}_{i,j} + \bm{B}_{i,j} \times \mathcal{CL}(0,1)$.
The third loss consists of the expectation over the auxiliary distributions of the log-likelihood of the received training signal. It can be derived in closed-form as in Eq. (\ref{eq:loss_3_laplace}), where $C_1$ is a constant, $\bm{Q}$ and $\bm{\Lambda}$ are the covariance matrix over the columns of $\bm{G}^\mathsf{vir}$ and covariance matrix of $\bm{h}^\mathsf{vir}$, respectively, which are diagonal matrices due to the independence of the elements according to the auxiliary distributions. The main diagonal elements are as follows (see the proof in the \textbf{Appendix}):
\begin{align}
    &\bm{\Lambda}_{i,i} = 6 \bm{b}_i^2; \; \bm{Q}_{i,i} = 6 \sum_{m=1}^M \bm{B}_{m,i}^2.
\end{align}

The parameters $\bm{m}$, $\bm{b}$, $\bm{M}$ and $\bm{B}$ of the auxiliary distributions are obtained using the variational neural networks, as shown in Eq. (\ref{eq:nn_eq}). Specifically, we employ \textit{Encoder} $\mathcal{F}$ to characterize $q_{\bm{\lambda}_1}(\bm{h}^\mathsf{vir}|\bm{Y})$ and \textit{Encoder} $\mathcal{G}$ for $q_{\bm{\lambda}_2}(\bm{G}^\mathsf{vir}|\bm{Y})$, i.e., $\bm{m}$ and $\bm{b}$ are the output of \textit{Encoder} $\mathcal{F}$ and $\bm{M}$ and $\bm{B}$ are the output of \textit{Encoder} $\mathcal{G}$. The training signal $\bm{Y}$ is fed to the encoders as input and the encoders' outputs are the parameters that maximize the ELBO. Given that the training signals involve complex numbers and neural networks typically operate with real-valued inputs, we preprocess the input by splitting it into its real and imaginary components. Subsequently, these components are concatenated before being fed into the neural networks. A similar approach is applied to the means of the auxiliary distributions. The output yields both the real and complex parts of the means, which are then used to reconstruct the complex numbers represented by $\bm{m}$ and $\bm{M}$.

\subsection{Joint Channel-Covariance Estimation via Variational Inference}

The JCE method estimates the UE-RIS and RIS-BS instantaneous channels separately in a fully passive RIS setting. However, we are interested in a solution with reduced training and signaling overheads. To this end, we extend this approach to exploit the slow-varying properties of the (i) RIS-BS channel as the physical locations of the RIS and BS do not change over time, and (ii) the UE-RIS CCM to perform the passive beamforming. We start by describing the transmission protocol and then introduce in detail the methodology and the ELBO derivations.

\subsubsection{Uplink Training}

\begin{figure}[!t]
    \centering
    \includegraphics[width=\linewidth]{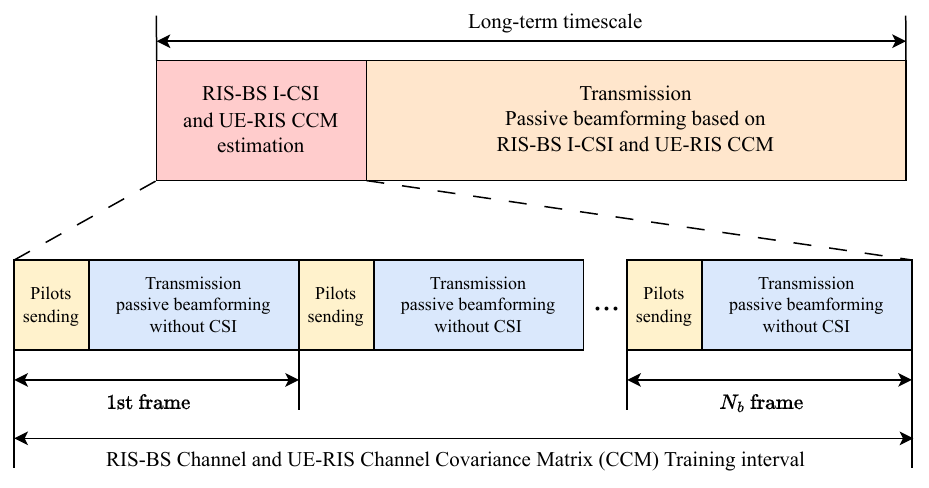}
    \caption{Transmission protocol.}
    \label{fig:transmission_protocol}
\end{figure}

We use the transmission protocol in Fig. \ref{fig:transmission_protocol} to effectively estimate the RIS-BS I-CSI and the UE-RIS CCM. Within the considered time interval, referred to as \textit{long-term timescale}, the UE-RIS channel varies according to the covariance matrix $\bb{E}[\bm{h}\bm{h}^{\mathsf{H}}]=\bm{R}_{\bm{h}}$ that remains invariant similar to the RIS-BS channel.
In alignment with the two-timescale training protocol outlined in \cite{wang2023covariance_estimation}, our approach involves a dual-phase process. In the initial phase, the focus is on estimating the RIS-BS I-CSI and the UE-RIS CCM. Then, the phase-shifts are optimized based on these estimates. Thus, the second phase is dedicated to transmissions where the optimized phase-shifts are fixed, and the channel estimation process focuses on estimating the $M \times 1$ low-dimension UE-RIS-BS effective channel alongside the data transmission.
Focusing on the initial phase, the considered interval is divided into $N_b$ coherence blocks of the UE-RIS channel $\bm{h}$ wherein we use the first $N_p$ time slots to send the training symbols, resulting in a total of $N_p \times N_b$ slots allocated for pilot transmission. To directly estimate $\bm{R}_{\bm{h}}$ from the training signal, it is essential that the training signal encompasses diverse realizations of $\bm{h}$. The remaining time slots in each coherence block of the channel $\bm{h}$ are then dedicated to transmissions, employing passive beamforming without CSI schemes such as in \cite{souto2020beamforming}.

In the $s$-th UE-RIS coherence block, by sending $N_p$ pilot signals while altering the configuration of each pilot, the received signal at the BS can be expressed as:
\begin{equation}
    \bm{Y}_s = \sqrt{\rho}\; \bm{G}\; \text{diag}(\bm{h}_s) \bm{\Phi} + \bm{W},\qquad {s=1,\dots,N_b},
\end{equation}
\noindent where $\bm{h}_s$ is the UE-RIS channel during the $s$-th coherence block, $\bm{\Phi}=[\bm{v}_1,\dots,\bm{v}_{N_p}] \in \mathbb{C}^{N\times N_p}$ is the RIS configuration used for training, $\bm{W}=[\bm{w}_1,\dots,\bm{w}_{N_p}]$ is the noise matrix where $\bm{w}_l \sim \mathcal{CN}(\bm{0},\bm{I}_N)$. The vectorized form of $\bm{Y}_s$ can be expressed as follows:
\begin{align}
    \bm{\tilde{y}}_s&=\text{vec}(\bm{Y}_s) =\sqrt{\rho} (\bm{\Phi}^T \odot \bm{G}) \bm{h}_s + \bm{w},
\end{align}
\noindent where $\bm{w}=\text{vec}(\bm{W}) \sim \mathcal{CN}(\bm{0},\bm{I}_{MN_p})$. We define the combined training received signal as $\bm{\tilde{Y}} = [\bm{\tilde{y}}_1,\dots,\bm{\tilde{y}}_{N_b}]$. The covariance matrix of the received training signal $\bm{\tilde{y}}_s$, given that the RIS-BS channel remains quasi-static, is expressed as:
\begin{equation}
    \bm{R}_{\bm{\tilde{y}}} = \mathbb{E}[\bm{\tilde{y}}_s\bm{\tilde{y}}_s^{\mathsf{H}}] = \rho (\bm{\Phi}^T \odot \bm{G})\bm{R}_{\bm{h}}(\bm{\Phi}^T \odot \bm{G})^{\mathsf{H}} + \bm{I}_{MN_p}.
\end{equation}
In various scenarios, the UE-RIS channel is highly correlated because of the small set of angles of arrivals (AoAs) contributing to the propagation \cite{haghighatshoar2016massive}. Therefore, the covariance matrix $\bm{R}_{\bm{h}}=\mathbb{E}[\bm{h}\bm{h}^{\mathsf{H}}]$ is considered as a low-rank matrix. Formally, we express the covariance matrix as follows:
\begin{equation}
    \bm{R}_{\bm{h}} = \bm{F}_N^{\mathsf{H}} \bm{D} \bm{F}_N,
\end{equation}
where $\bm{D}=\text{diag}(\bm{d})$ is a diagonal matrix with a sparse main diagonal denoted as $\bm{d}$. We focus on estimating the sparse vector $\bm{d}$, rather than estimating the full covariance matrix $\bm{R}_{\bm{h}}$ which is typically a large matrix of size $N \times N$. For the RIS-BS channel, we use the following representation in the angular domain: $\bm{G}^\mathsf{vir}= \bm{F}_M \bm{G} \bm{F}_N$.

\subsubsection{Derivation of the ELBO}

As discussed in the previous subsection, the channel between the RIS and the BS exhibits sparsity in the angular domain. The complex Laplace distribution is employed to model the sparse matrix $\bm{G}^\mathsf{vir}$. Additionally, the vector $\bm{d}$, which represents a sparse positive real-valued vector, is modeled using a complex Exponential distribution:
\begin{align}
    &\bm{G}^\mathsf{vir}_{i,j} \sim \mathcal{CL}(0, \alpha_{\bm{G}^\mathsf{vir}});\; \bm{d}_i \sim \text{Exp}(\alpha_{\bm{d}}).
\end{align}

Applying the VI framework, we approximate the intractable true posterior distribution $p(\bm{G}^\mathsf{vir},\bm{d}|\bm{\tilde{Y}})$ by two separate tractable parameterized distributions denoted by $q_{\bm{\lambda}_1}(\bm{d}|\bm{\tilde{Y}})$ and $q_{\bm{\lambda}_2}(\bm{G}^\mathsf{vir}|\bm{\tilde{Y}})$ using the mean-field approximation. Moreover, the parameters of the chosen auxiliary distributions are returned by \textit{Encoder} $\mathcal{F}$ and \textit{Encoder} $\mathcal{G}$. The training signal $\bm{\tilde{Y}}$ is preprocessed such that the input to the neural networks is defined by $\bm{\tilde{Y}} \bm{\tilde{Y}}^{\mathsf{H}} / N_b - \bm{I}_{MN_p}$.

The auxiliary distribution for the RIS-BS channel in the angular domain $\bm{G}^\mathsf{vir}$ is assumed to follow the complex Laplace distribution with independent elements and the elements of $\bm{d}$ follow a Gamma distribution with unit scale:
\begin{align}
    &q_{\bm{\lambda}_1}(\bm{d}_{i}|\bm{\tilde{Y}}) \sim \text{Gamma}(\bm{k}_i);\\
    &q_{\bm{\lambda}_2}(\bm{G}^\mathsf{vir}_{i,j}|\bm{\tilde{Y}}) \sim \mathcal{CL}(\bm{M}_{i,j},\bm{B}_{i,j}),
\end{align}
where $\bm{\lambda_1}=\{\bm{k}\}$ and $\bm{\lambda_2}=\{\bm{M},\bm{B}\}$ are the parameters of the auxiliary distributions which are obtained by minimizing the following loss function:
{\small
\begin{align}\label{eq:elbo_jcce}\scriptsize
    \mathcal{L}^\mathsf{S-CSI}({\bm{\lambda}}_1,{\bm{\lambda}}_2)=& \underbrace{\mathbb{E}_{\bm{d} \sim q_{\bm{\lambda}_1}(\bm{d}|\bm{\tilde{Y}})} \left[\log \frac{q_{\bm{\lambda}_1}(\bm{d}|\bm{\tilde{Y}})} {p(\bm{d})} \right]}_{\mathcal{L}^\mathsf{S-CSI}_1}\nonumber\\
    &+ \underbrace{\mathbb{E}_{\bm{G}^\mathsf{vir} \sim q_{\bm{\lambda}_2}(\bm{G}^\mathsf{vir}|\bm{\tilde{Y}})} \left[\log \frac{q_{\bm{\lambda}_2}(\bm{G}^\mathsf{vir}|\bm{\tilde{Y}})} {p(\bm{G}^\mathsf{vir})} \right]}_{\mathcal{L}^\mathsf{S-CSI}_2} \nonumber\\
    &\underbrace{- \mathbb{E}_{\bm{d},\bm{G}^\mathsf{vir} \sim q_{\bm{\lambda}}(\bm{d},\bm{G}^\mathsf{vir}|\bm{\tilde{Y}})} \bigl[\log p(\bm{\tilde{Y}}|\bm{d},\bm{G}^\mathsf{vir})\bigl]}_{\mathcal{L}^\mathsf{S-CSI}_3}.
\end{align}
}
\normalsize
The expression of the second loss function $\mathcal{L}^\mathsf{S-CSI}_2$ is the same as $\mathcal{L}^\mathsf{I-CSI}_2$ in Eq.(~\ref{eq:l2_jce}) since the prior and the auxiliary posterior of $\bm{G}^\mathsf{vir}$ are the same.
The first loss, which involves the KL-divergence between an Exponential distribution and a Gamma distribution, can be expressed as follows:
\small
\begin{align}
    \mathcal{L}_1^\mathsf{S-CSI}({\bm{\lambda}_1}) &= \mathbb{E}_{\bm{d} \sim q_{\bm{\lambda}_1}(\bm{d}|\bm{\tilde{Y}})} \left[\log \frac{q_{\bm{\lambda}_1}(\bm{d}|\bm{\tilde{Y}})} {p(\bm{d})} \right]\nonumber\\
    &=\sum_{i=1}^N \mathbb{E}_{\bm{d}_i \sim q_{\bm{\lambda}_1}(\bm{d}_i|\bm{\tilde{Y}})} \left[\log q(\bm{d}_i|\bm{\tilde{Y}}) - \log {p(\bm{d}_i)} \right]\nonumber\\
    &=\sum_{i=1}^N (1- \bm{k}_i) \psi(1) - \log \Gamma(1.0) + \log \Gamma(\bm{k}_i),
\end{align}\normalsize
where $\Gamma(x)$ is the gamma function and $\psi(x)$ is the digamma function. 
The third loss $\mathcal{L}_3^\mathsf{S-CSI}$ is defined as the log-likelihood of the received training signal and can be expressed as follows:
\small
\begin{align}
\label{eq:L3_SCSI}
    \mathcal{L}_3^\mathsf{S-CSI}({\bm{\lambda}_1},{\bm{\lambda}_2})
    &=\mathbb{E}_{\bm{d},\bm{G}^\mathsf{vir} \sim q_{\bm{\lambda}}(\bm{d},\bm{G}^\mathsf{vir}|\bm{\tilde{Y}})} \Bigr[\trace\left(\bm{\tilde{Y}}^{\mathsf{H}} \bm{R}_{\bm{\tilde{Y}}}^{-1} \bm{\tilde{Y}}\right) \nonumber\\
    &\quad+ \log \left|\bm{R}_{\bm{\tilde{Y}}}\right|\Bigr] + C_2,
\end{align}\normalsize
where $C_2$ is a constant.
To compute the gradient with respect to the parameters of the auxiliary distribution of the RIS-BS channel link, $q_{\bm{\lambda}_2}(\bm{G}^\mathsf{vir}|\bm{\tilde{Y}})$, we use the reparameterization trick where the Monte-Carlo samples are computed by $\widehat{\bm{G}^\mathsf{vir}}_{i,j}=\bm{M}_{i,j} + \bm{B}_{i,j}\times \mathcal{CL}(0,1)$. Applying the reparameterization trick for the Gamma distribution is less straightforward. Hence, we use an alternative technique known as the implicit reparameterization \cite{figurnov2018implicit} which facilitates the generation of Monte-Carlo samples that remain differentiable with respect to the shape parameter vector $\bm{k}$.

After training the neural networks, \textit{Encoder} $\mathcal{F}$ and \textit{Encoder} $\mathcal{G}$, that predict the distribution parameters $\bm{k}$ and $\{\bm{M}, \bm{B}\}$ of $q_{\bm{\lambda}_1}(\bm{d}|\bm{\tilde{Y}})$ and $q_{\bm{\lambda}_2}(\bm{G}^\mathsf{vir}|\bm{\tilde{Y}})$, respectively, the channels are estimated using the MAP method applied on the auxiliary distributions:
\begin{align}
    &\widehat{\bm{d}} = \arg \max_{\bm{d}}\; q_{\bm{\lambda}_1}(\bm{d}|\bm{\tilde{Y}}) = \bm{k} - \bm{1};\\
    &\widehat{\bm{G}^\mathsf{vir}} = \arg \max_{\bm{G}^\mathsf{vir}} \;q_{\bm{\lambda}_2}(\bm{G}^\mathsf{vir}|\bm{\tilde{Y}}) = \bm{M}.
\end{align}

\section{Optimization of RIS Phase-shifts}

The primary evaluation metric is the capacity of the RIS-assisted network obtained after deriving the phase-shifts based on the estimated quantities. Therefore, we derive closed-form expressions of the phase-shifts of the RIS that maximize the capacities for the two types of channels considered in the two proposed solutions JCE and JCCE.

\subsection{Instantaneous CSI}

The ergodic capacity of the uplink RIS-assisted mmWave system is given by:
\begin{equation}
    C = \log_2\bigl(1+\rho\; ||\bm{G}\; \diag(\bm{v})\;\bm{h}||^2_2\bigl).
\end{equation}
Based on the I-CSI (i.e., $\bm{h}$ and $\bm{G}$), we configure the phase-shifts to maximize the capacity $C$, which is equivalent to solving the following problem:
\begin{equation}
\begin{split}
    &\max_{\{\theta_i\}} \;||\bm{G}\; \diag(\bm{v})\; \bm{h}||^2_2\\
    &\text{Subject to: } \bm{v}_{i}=e^{j\theta_i}, \qquad  i = 1,\dots,N.
\end{split}
\end{equation}
Given the singular value decomposition (SVD) of $\bm{G}=\bm{U}\bm{S}\bm{V}^{\mathsf{H}}$, the problem is equivalent to maximizing $||\bm{SV^{\mathsf{H}}}\;\diag(\bm{v})\; \bm{h}||^2_2$ which is expressed as follows:
\begin{align}
    ||\bm{S}\bm{V}^{\mathsf{H}}\diag(\bm{v}) \bm{h}||^2_2 &= \sum_{i=1}^r \left|\sum_{k=1}^N \zeta_i \bm{V}_{ki}^* \bm{h}_k \bm{v}_k\right|^2 \nonumber \\
    &=\sum_{i=1}^r \left|\sum_{k=1}^N \zeta_i |\bm{V}_{ki}| |\bm{h}_k| e^{j(\theta_k - \angle \bm{V}_{ki} + \angle \bm{h}_k)}\right|^2,
\end{align}
where $r$ is the rank of $\bm{G}$ and $\zeta_i$ are the singular values in the descending order of $\bm{G}$.
The solution we propose is to align the phase-shifts ${\theta}_k$ to the phases of the largest right singular vector of $\bm{G}$, denoted as $\bm{\vartheta}^{\max}$, and the phases of the channel vector $\bm{h}$.
Specifically, the suboptimal phase-shifts are obtained as follows:
\begin{equation}
    \theta_k^* = -(\angle \bm{h}_{k} -  \angle \bm{\vartheta}^{\max}_k).
\end{equation}

\subsection{RIS-BS I-CSI and UE-RIS S-CSI}
In this section, we propose a closed-form expression of the phase-shifts that maximize the achievable rate of the UE-RIS-BS link based on the I-CSI of RIS-BS channel and the S-CSI (i.e., channel covariance matrix) of the UE-RIS channel. The problem is formulated as follows:
\begin{equation}
\label{eq:capacity_I-CSI}
\begin{split}
    &\max_{\{\theta_i\}}\; \bb{E}_{\bm{h}}\Bigr[\log_2 \left(1 + \rho \;||\bm{G} \;\diag(\bm{v})\; \bm{h}||^2_2\right)\Bigr],\\
    &\text{Subject to: } \bm{v}_i=e^{j\theta_i}, \qquad i=1,\dots,N.
\end{split}
\end{equation}
The problem in Eq.~(\ref{eq:capacity_I-CSI}) is challenging to solve due to the lack of an explicit expression of the expectation over the logarithm. To address this difficulty, we adopt a strategy of maximizing a reliable upper bound on this expression \cite{zhao2020tts}:

{\footnotesize\begin{equation}
\label{eq:capacity_upper_bound}
    \bb{E}_{\bm{h}}\left[\log_2 \left(1 + \rho||\bm{G} \diag(\bm{v}) \bm{h}||^2_2\right)\right] \leq \log_2 \left(1 + \rho \bb{E}_{\bm{h}}\left[||\bm{G} \diag(\bm{v}) \bm{h}||^2_2\right]\right).
\end{equation}
}It is important to acknowledge that the upper bound in  Eq.~(\ref{eq:capacity_upper_bound}) is highly accurate and serves as a reliable approximation of the original objective function, particularly for large values of $\rho$ \cite{zhao2020tts}. To maximize this upper bound, we can formulate the subsequent optimization problem as follows:
\begin{equation}
\begin{split}
    &\max_{\{\theta_i\}}\; \bb{E}_{\bm{h}} \Bigr[||\bm{G}\; \diag(\bm{v})\; \bm{h}||_2^2\Bigr],\\
    &\text{Subject to: } \bm{v}_i = e^{j\theta_i}, \qquad i=1,\dots N.
\end{split}
\end{equation}
The objective can be further expressed as follows:
\small
\begin{align}
    \bb{E}_{\bm{h}}\Bigr[||\bm{G} \diag(\bm{v}) \bm{h}||^2_2\Bigr]
    =\trace\Bigl(\bm{G}\diag(\bm{v})\bm{R}_{\bm{h}} \diag(\bm{v})^{\mathsf{H}} \bm{G}^{\mathsf{H}}\Bigl).
\end{align}\normalsize
Given the SVD of $\bm{G}=\bm{U}\bm{S}\bm{V}^{\mathsf{H}}$ and the eigenvalue decomposition of the covariance matrix $\bm{R}_{\bm{h}} = \bm{P} \bm{\Sigma} \bm{P}^{\mathsf{H}}$, the objective function can be expressed as follows:
\begin{align}
    \bb{E}_{\bm{h}}\Bigr[||\bm{G} \diag(\bm{v}) \bm{h}||^2_2 \Bigr]&= \sum_{i=1}^r \sum_{j=1}^{r'} \left|s_i\sqrt{\sigma_j} \sum_{k=1}^N \bm{V}^*_{k,i} \bm{P}_{k,j}e^{j\theta_k} \right|^2,
\end{align}
where $r'$ is the rank of $\bm{R}_{\bm{h}}$ and $\sigma_j$ are the eigenvalues of $\bm{R}_{\bm{h}}$ in the descending order.
Therefore, we take the phases that align with the phases of the largest eigenvector of $\bm{G}$ and $\bm{R}_{\bm{h}}$, referred to as $\bm{\vartheta}^{\max}$ and $\bm{p}^{\max}$, respectively, to maximize the objective function and satisfy the unit modulus constraints, which are given by:
\begin{equation}
    \theta_k^* = -(\angle \bm{p}^{\max}_k - \angle \bm{\vartheta}^{\max}_k).
\end{equation}

\section{Simulation Setup}

In this section, we evaluate the performance of the two proposed CSI estimation methods in RIS-aided SIMO mmWave wireless communication systems. We consider the setup of $M=4$ antennas at the BS and $N=64$ passive elements at the RIS.

\subsection{Evaluation Metrics and Baselines}

The primary evaluation metric is the capacity of the RIS-aided SIMO communication system. 
Moreover, we evaluate the normalized mean square error (NMSE) defined by $\text{NMSE}=||\hat{\bm{X}} - \bm{X}||^2/||\bm{X}||^2_2$, where Frobenius norm is used for matrices and $l_2$ norm is used for vectors.

We compare our approaches against the following baselines:
\begin{itemize}
    \item \textbf{Perfect CSI}: this is an upper bound where the capacity is obtained based on the optimal phase shifts using the true channels $\bm{G}$ and $\bm{h}$;
    \item \textbf{Perfect channel and perfect covariance (PC-PCov)}: the capacity is computed based on the true RIS-BS channel $\bm{G}$ and the true UE-RIS CCM $\bm{R}_{\bm{h}}$;
    \item \textbf{Random phase-shifts}: this represents a lower bound for our method.
    \item \textbf{MO-EST} \cite{lin2022mo_est}: This method is based on alternating minimization and manifold optimization.
\end{itemize}

\begin{figure*}[!t]
    \centering
    \begin{subfigure}{.49\textwidth}
        \centering
        \includegraphics[width=0.8\linewidth]{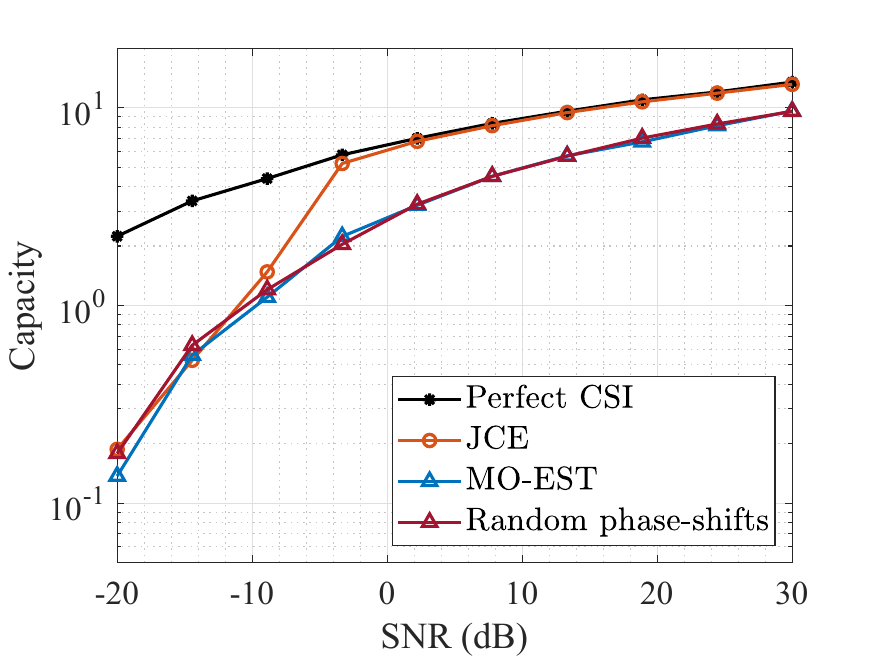}
        \caption{Achieved capacity}
        \label{fig:cap_jce}
    \end{subfigure}
    \begin{subfigure}{.49\textwidth}
        \centering
        \includegraphics[width=0.8\linewidth]{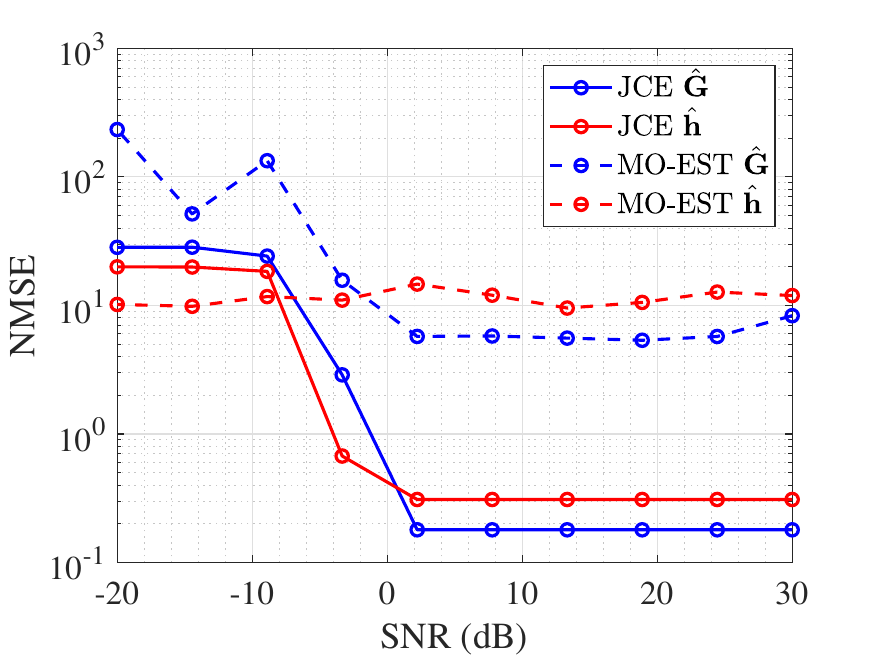}
        \caption{NMSE}
        \label{fig:nmse_jce}
    \end{subfigure}
    \caption{Performance of JCE method.}
    \label{fig:perf_jce}
\end{figure*}
\begin{figure*}[!t]
    \centering
    \begin{subfigure}{.49\textwidth}
        \centering
        \includegraphics[width=0.8\linewidth]{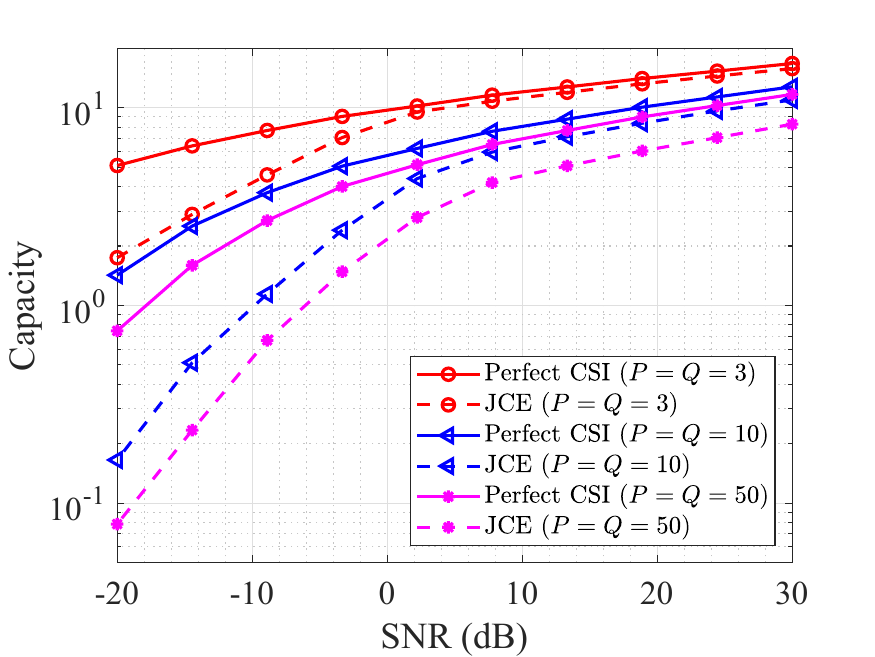}
        \caption{Achieved capacity}
        \label{fig:cap_npaths}
    \end{subfigure}
    \begin{subfigure}{.49\textwidth}
        \centering
        \includegraphics[width=0.8\linewidth]{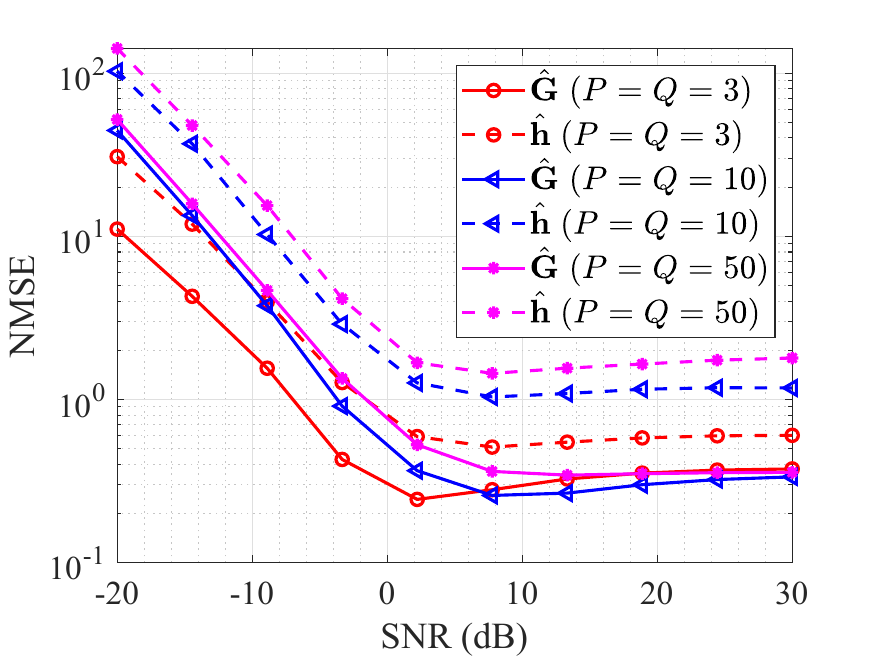}
        \caption{NMSE}
        \label{fig:nmse_npaths}
    \end{subfigure}
    \caption{Performance of JCE with different number of paths.}
    \label{fig:perf_npaths}
\end{figure*}

\subsection{Model Details and Hyperparameter Settings}
We conduct an exhaustive hyperparameter search to select the encoder architectures and the training hyperparameters.
The hyperparameter tuning is conducted using Bayesian optimization \cite{bayesian_hyper_tune}. 
The hyperparameters tuned consist of the architecture of the neural networks, the use of dropout layers, and the learning rate.
The architecture adopted for the JCE method features fully connected neural networks for both \textit{Encoder} $\mathcal{F}$ and \textit{Encoder} $\mathcal{G}$. They consist of an input layer, two 300-unit hidden layers with Relu activation combined with a dropout layer and a batch normalization layer, and an output layer with two heads: the first outputs the mean after a Tanh activation and the second uses Softmax activation for scale.
Conversely, for the JCCE method, we maintain the architecture of the encoders and we adapt the output layer of \textit{Encoder} $\mathcal{F}$ to consist of one head with a Sigmoid activation modeling the auxiliary distribution $q_{\bm{\lambda}_1}(\bm{d}|\bm{\tilde{Y}})$.
Adam optimizer \cite{kingma2014adam} is used to train the neural networks with $0.1$ as an initial learning rate. The neural networks are trained by maximizing the ELBO functions using $10^4$ unlabeled samples. The priors' statistical parameters are chosen as $\alpha_{\bm{G}^\mathsf{vir}}=\alpha_{\bm{h}^\mathsf{vir}}=\alpha_{\bm{d}}=1$. The expectations within the objective functions, i.e., Eq. (\ref{eq:L1_ICSI}), Eq. (\ref{eq:l2_jce}), and Eq. (\ref{eq:L3_SCSI}), are evaluated using Monte-Carlo with $1000$ samples.
The methods are tested based on $50$ Monte-Carlo samples.

\subsection{Channel Model}
We adopt a mmWave channel model as follows \cite{lin2022mo_est}:
\begin{align}
    &\bm{G} = \sqrt{\frac{MN}{P}} \sum_{p=1}^P \alpha_p \bm{a}_{\BS}(\xi_p) \bm{a}_{\RIS}^{\mathsf{H}}(\phi_p,\varphi_p);\\
    &\bm{h} = \sqrt{\frac{N}{Q}} \sum_{q=1}^Q \beta_q \bm{a}_{\RIS}(\phi_q, \varphi_q),
\end{align}
where $\alpha_p$, $\xi_p$, and $\phi_p$/$\varphi_p$ denote the complex gain, AoA, and azimuth/elevation AoD of the $p$-th path of RIS-BS channel. Similarly, $\beta_q$ and $\phi_q$/$\varphi_q$ denote the complex gain and azimuth/elevation AoA of the $q$-th path of the UE-RIS channel, respectively. Besides, $\bm{a}_{\BS}$ and $\bm{a}_{\RIS}$ denote the receive and transmit array response vectors at the BS and the RIS, respectively. Then, the array response vector of the half-wavelength spaced uniform linear array at the BS is given by:
\begin{align}
    \bm{a}_{\BS}(\xi_p) = \frac{1}{\sqrt{M}} \left[1, e^{j\pi \cos\xi_p}, \dots, e^{j\pi (M-1) \cos \xi_p}\right]^T.
\end{align}
In addition, the array response vector of the planar array at the RIS involving $N$ elements is given by:
\begin{equation}
    \bm{a}_{\RIS}(\phi,\varphi) = \frac{1}{\sqrt{N}}\begin{bmatrix}
        1 \\
        e^{j\pi \sin\phi \sin\varphi}\\
        \vdots \\
        e^{j\pi \sqrt{N} \sin\phi \sin\varphi}
    \end{bmatrix}
    \otimes
    \begin{bmatrix}
        1 \\
        e^{j\pi \cos\varphi} \\
        \vdots \\
        e^{j\pi \sqrt{N} \cos\varphi}
    \end{bmatrix}.
\end{equation}
We distinguish two channel generation modes to train and test our methods: 
\begin{itemize}
    \item \textbf{Mode 1}: The AoAs $\phi_q$ and $\varphi_q$ are uniformly generated from the interval $[0, 2\pi)$, and this mode is used to evaluate the JCE method.
    \item \textbf{Mode 2}: It adopts a different approach by generating AoAs $\phi_q$ and $\varphi_q$ from different clusters, dividing the interval $[0, 2\pi)$ into $100$ sub-intervals. This clustering results in a covariance matrix that exhibits sparsity in the angular domain. This mode is used to evaluate the JCCE approach.
\end{itemize}

\section{Simulation Results}
\subsection{Performance of JCE}

We evaluate the performance of the proposed JCE method using mmWave channels generated according to Mode 1. To estimate the UE-RIS and the RIS-BS channels, we send $N_p = 50$ pilot symbols over an uplink SIMO RIS-assisted mmWave communication system with number of paths $Q=1$ and $P=3$ for the UE-RIS and RIS-BS channels, respectively, and obtain the training signals which are fed to the trained neural networks \textit{Encoder} $\mathcal{F}$ and \textit{Encoder} $\mathcal{G}$.
Fig. \ref{fig:cap_jce} illustrates the capacity as a function of the SNR $\rho$. 
The phase-shifts derived from the estimated channels achieve a better capacity than the random selection of the RIS configuration which validates that the neural networks can effectively learn the channels. Moreover, our method outperforms the MO-EST method primarily due to the ability of neural networks to capture the sparse structure of the channels at high dimensions. In particular, the JCE method demonstrates a notable improvement with a gain of 3.70 dB at -3.33 dB SNR compared to the MO-EST method and achieves a gain of 1.35 dB at 30 dB SNR.

Next, we investigate the estimation error of both channels UE-RIS and RIS-BS. As depicted in Fig. \ref{fig:nmse_jce}, the NMSE decreases as the SNR increases. Notably, our learning-based approach significantly outperforms the MO-EST baseline.
In addition, the proposed method presents a lower computation time than the iterative algorithm MO-EST by leveraging the significantly lower inference time of the neural networks. Specifically, at 20dB of SNR, the neural networks predict the auxiliary parameters within 0.20 seconds, whereas MO-EST requires 1.45 seconds to estimate the channels.

Furthermore, we evaluate the JCE method under a different number of paths investigating the effect of the level of sparsity on the estimation performance. Fig. \ref{fig:cap_npaths} presents the capacity as a function of the SNR for three scenarios: $P=Q=3$, $P=Q=10$, and $P=Q=50$. The numerical results reveal that, under high SNR, the capacity achieved based on phase-shifts derived from the estimated channels converges towards the exact capacity obtained when employing phase-shifts derived from the perfect CSI. Furthermore, we observe a notable impact of channel sparsity on the estimation performance in terms of capacity. Specifically, as the channel sparsity increases, signifying a reduced number of propagation paths, the achieved capacity becomes increasingly closer to the exact capacity due to the improvement of estimation of the channels. This behavior can be attributed to the sparsity-inducing nature of the variational loss function employed by the encoders, which leverages a Laplace prior to enforcing a sparse structure over the channels. Consequently, the proposed JCE method demonstrates superior performance for scenarios involving more sparse channels compared to those with less sparsity.
\begin{figure}[!t]
    \centering
    \includegraphics[width=0.8\linewidth]{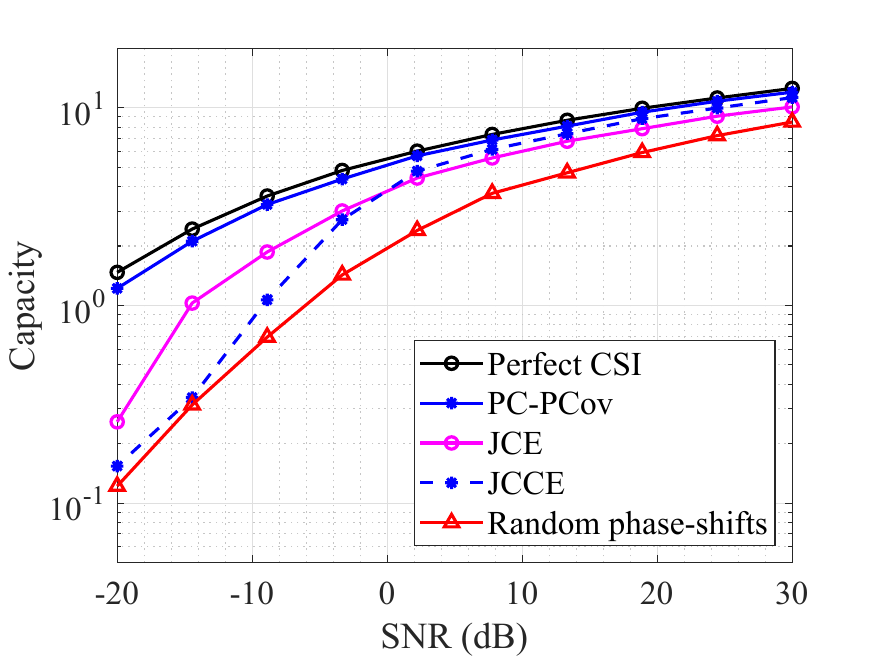}
    \caption{Performance of the proposed methods.}
    \label{fig:perf_comp}
\end{figure}
Fig. \ref{fig:nmse_npaths} depicts the evaluation of the NMSE to assess the performance of the proposed method. Notably, as the SNR increases, a clear trend emerges where the NMSE consistently decreases. Additionally, the degree of sparsity in the channel in the angular domain $\bm{h}^\mathsf{vir}$ plays a critical role \cite{xu2019secure}. More specifically, the NMSE exhibits a significant degradation when the number of paths increases, with the most substantial performance deterioration occurring when $P=Q=50$ paths are considered. This degradation can be attributed to the fact that $50$ paths approach the dimensionality of the channel vector $\bm{h}\in \bb{C}^{64}$. Conversely, for the RIS-BS channel $\bm{G}^\mathsf{vir}$, the NMSE experiences a minor degradation as the number of paths varies. This behavior stems from the larger dimensionality of the RIS-BS channel matrix, $M\times N = 256$, in relation to the maximum number of paths, mitigating the impact of variations in the number of paths. Importantly, these findings highlight the superior efficiency of the proposed method in scenarios characterized by higher levels of sparsity, effectively bypassing the need for a priori knowledge of the specific number of paths.

\begin{figure*}[!t]
    \centering
    \begin{subfigure}[t]{.49\textwidth}
        \centering
        \includegraphics[width=0.8\linewidth]{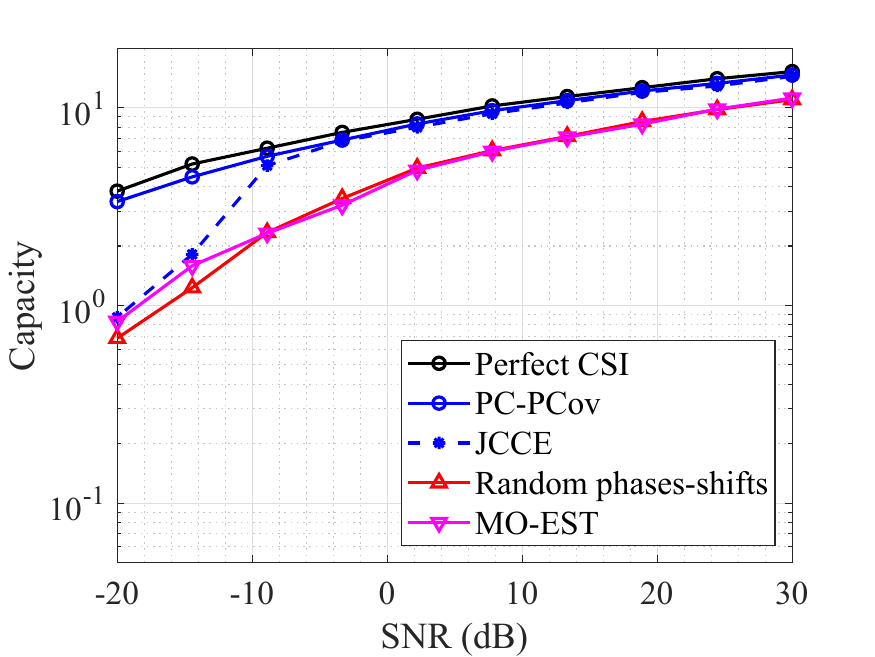}
        \caption{Achieved capacity}
        \label{fig:capacity_result1}
    \end{subfigure}
    \begin{subfigure}[t]{.49\textwidth}
        \centering
        \includegraphics[width=0.8\linewidth]{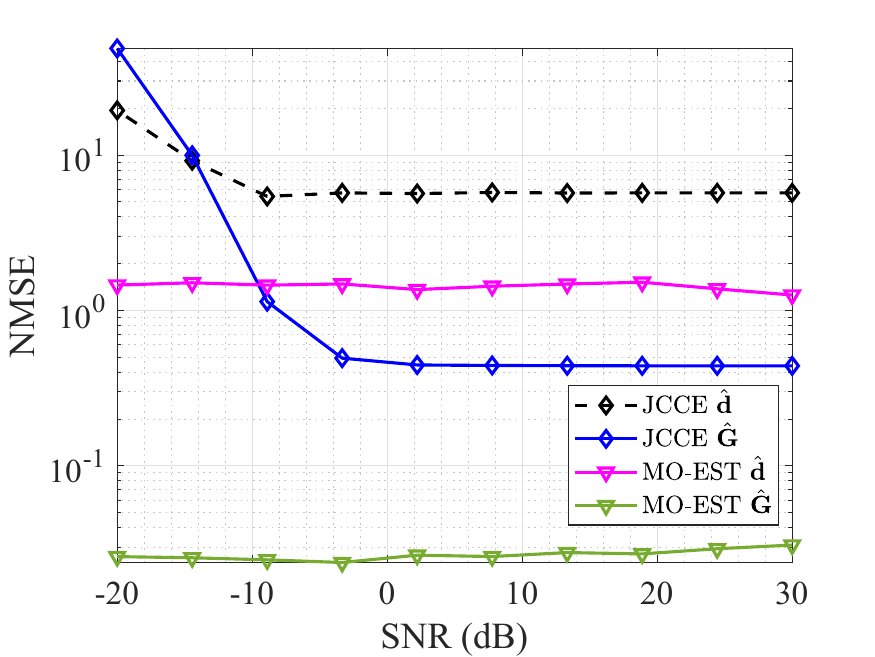}
        \caption{NMSE}
        \label{fig:nmse_result1}
    \end{subfigure}
    \caption{Performance of the VI-based estimation of RIS-BS channel and UE-RIS channel covariance matrix.}
    \label{fig:perf_vi2}
\end{figure*}

\begin{figure}[!t]
    \centering
    \includegraphics[width=0.8\linewidth]{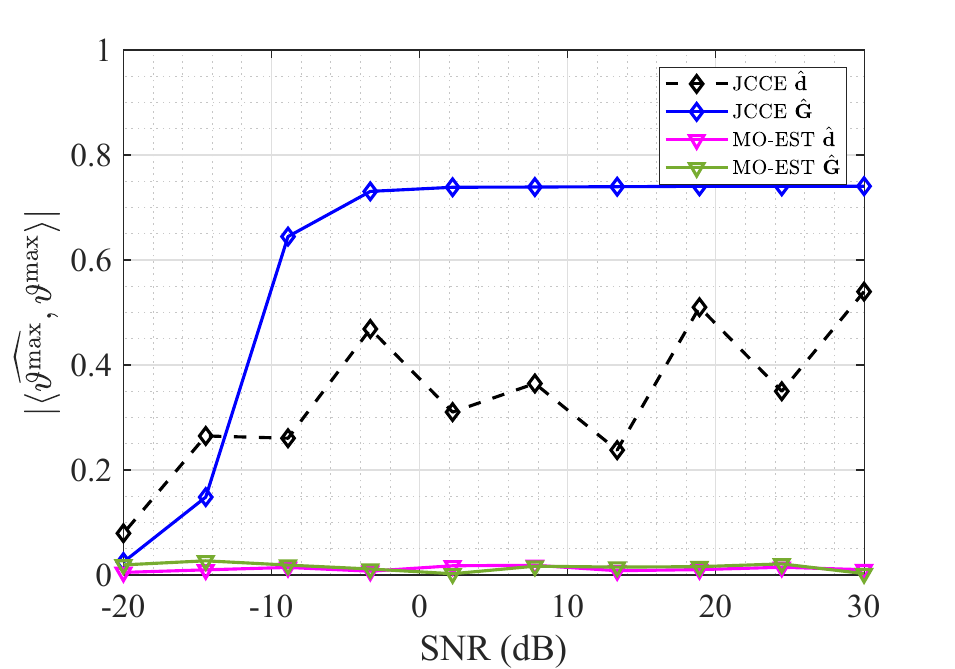}
    \caption{Inner Product of the estimated largest eigenvectors with ground truth.}
    \label{fig:inner_prod}
\end{figure}

\subsection{Performance of Joint Channel-Covariance Estimation}

For our simulations, we select the following parameter values: $N_p=4$ for the number of pilot symbols per UE-RIS coherence block and $N_b=200$ for the number of coherence blocks for UE-RIS channel. To evaluate the JCCE method, we compare it against the MO-EST estimation approach, where the channels are estimated at each coherence block and used to estimate the covariance matrix $\bm{R}_{\bm{h}}$. We set $P=3$ and $Q=1$ to represent the number of paths for the RIS-BS and UE-RIS channels, respectively.
\begin{figure}[!ht]
    \centering
    \includegraphics[width=0.8\linewidth]{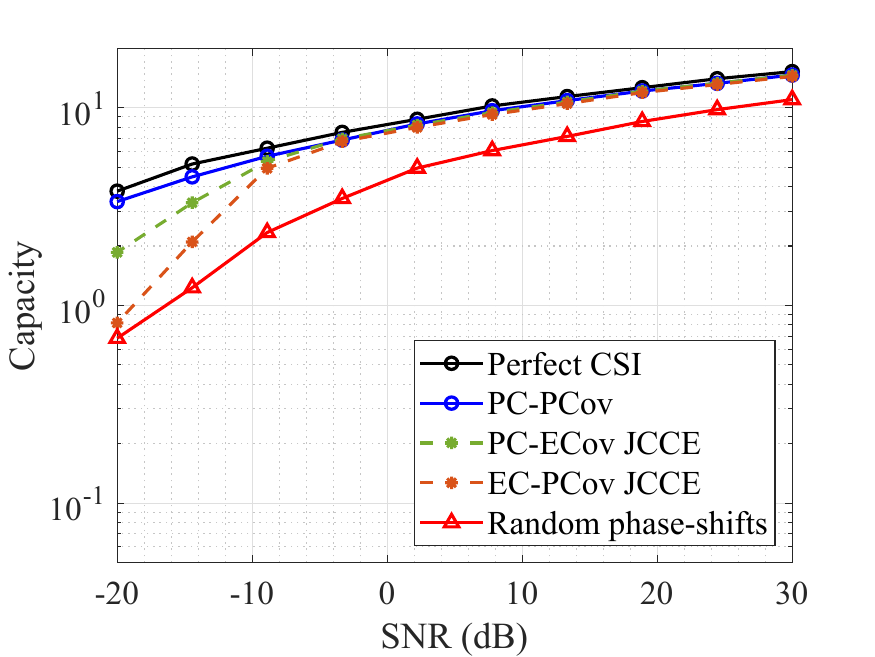}
    \caption{Performance of the estimates separated.}
    \label{fig:perf_sep}
\end{figure}
Fig. \ref{fig:capacity_result1} shows a degradation in performance by substituting the UE-RIS covariance matrix (PC-PCov) for the UE-RIS channel itself (Perfect CSI). However, by updating the RIS phase-shifts based on the UE-RIS CCM, we reduce the signaling overhead associated with the RIS configuration. This approach enables the RIS configuration to remain fixed for an extended period while ensuring an acceptable rate performance since the UE-RIS CCM and the RIS-BS channel are considered quasi-static for the subsequent coherence blocks of the UE-RIS channel.
Moreover, the capacity values using the phase-shifts derived from the estimated channel and the CCM via JCCE get closer with the increase of the SNR to the exact capacity which validates the proposed method. Furthermore, the proposed method demonstrates superior performance compared to the MO-EST method which fails to capture the sparse structure of the channel and its covariance.
\begin{figure*}[!ht]
    \centering
    \begin{subfigure}{.49\textwidth}
        \centering
        \includegraphics[width=0.8\linewidth]{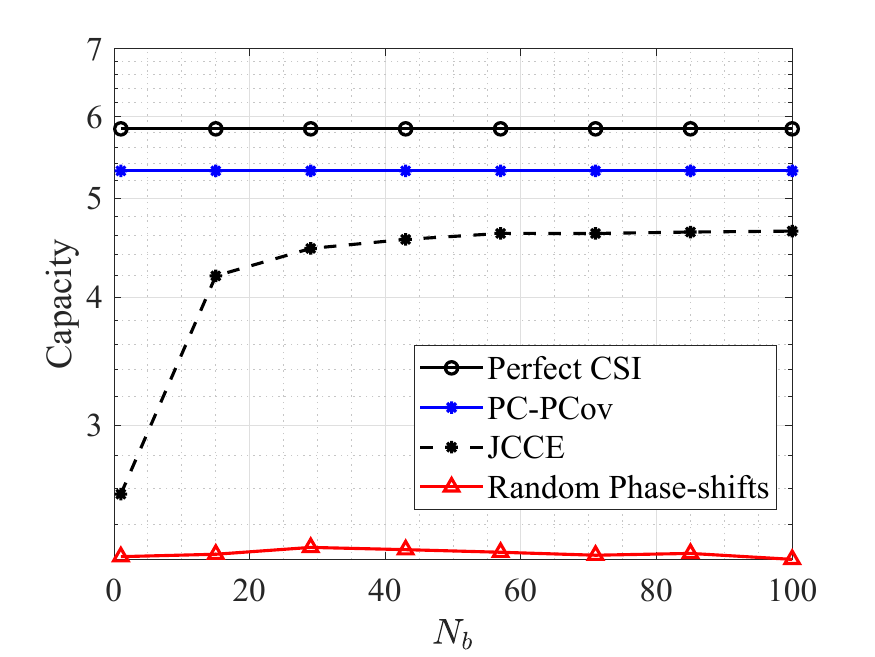}
        \caption{Achieved capacity}
        \label{fig:cap_jcce_coh}
    \end{subfigure}
    \begin{subfigure}{.49\textwidth}
        \centering
        \includegraphics[width=.8\linewidth]{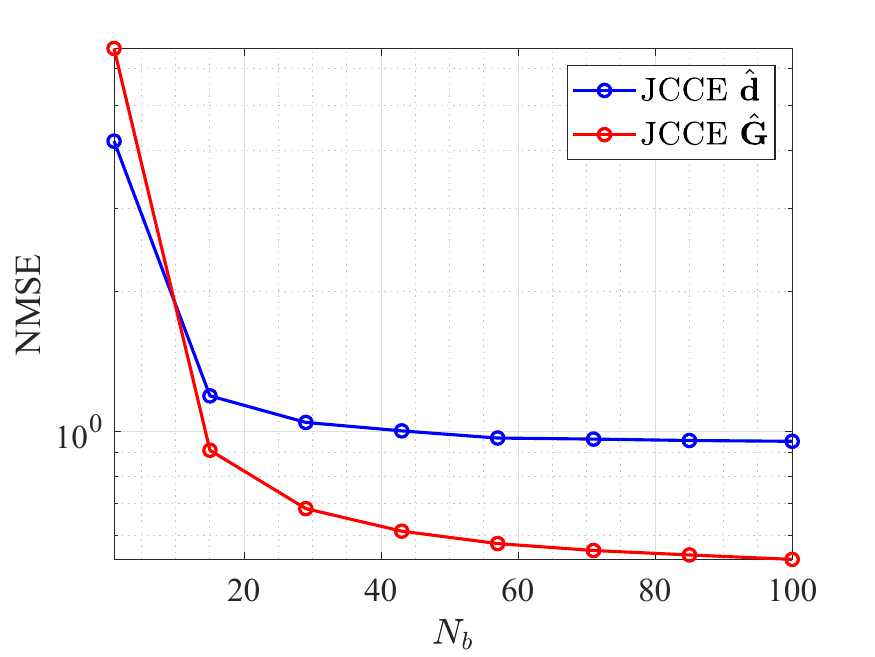}
        \caption{NMSE}
        \label{fig:nmse_jcce_coh}
    \end{subfigure}
    \caption{Performance of JCCE \textit{v.s} the number coherence blocks}
    \label{fig:jcce_coh}
\end{figure*}
Fig. \ref{fig:nmse_result1} showcases the NMSE evaluation across different SNR values. Notably, the MO-EST method reaches lower values of NMSE compared to the proposed method for the RIS-BS channel $\bm{G}$ and the angular spectrum $\bm{d}$. For further investigation, in Fig. \ref{fig:inner_prod}, we evaluate the absolute value of the complex inner product of the largest eigenvectors of the estimated RIS-BS channels $\widehat{\bm{G}}$ and the estimated CCM $\widehat{\bm{R}}_{\bm{h}}$, expressed as $\langle \widehat{{\bm{\vartheta}}^{\max}}, \bm{\vartheta}^{\max}\rangle= \widehat{{\bm{\vartheta}}^{\max}}^{\mathsf{H}} \bm{\vartheta}^{\max}$, with the largest eigenvectors from the PC-PCov. We observe that the proposed method is able to effectively estimate the largest eigenvectors of the RIS-BS channel and the UE-RIS covariance matrix, as the inner product gets closer to 1, with the increase of the SNR. This can be interpreted as an alignment of the estimated largest eigenvector to the largest eigenvector of the actual channel and covariance matrix.

\subsection{Comparison Between the Proposed Methods}

To compare the JCE and JCCE methods, we evaluate the capacity considering the training overhead that is expressed as $C_p = (1-\alpha) \log_2(1+\rho ||\bm{G}\diag(\bm{v})\bm{h}||^2)$ where $\alpha=N_{\text{pilot transmissions}}/N_{\text{Total transmissions}}$. We consider the parameters $N_p = 4$ and $N_b=200$ to obtain the training signal with channels generated in Mode 2.
At coherence times $T_{\bm{G}}$ and $T_{\bm{h}}$ in the order of $100$ms and $0.1$ms, respectively, Fig. \ref{fig:perf_comp} shows that the JCE method exhibits superior performance over the JCCE approach at low SNR, while the JCCE method outperforms the JCE at high SNR when the estimates closely approach the PC-PCov. The observed performance improvement can be attributed to the inherent differences in their channel estimation approaches. With the JCE, the channels $\bm{G}$ and $\bm{h}$ need to be estimated at each coherence block of $\bm{h}$, which is relatively short compared to the quasi-static nature of channel $\bm{G}$ and the covariance $\bm{R}_{\bm{h}}$, and thus it leads to higher values of $\alpha$, i.e., higher training overhead. In contrast, the JCCE method utilizes the estimates of the RIS-BS channel and the UE-RIS CCM, enabling the use of phase-shifts without the need to estimate the UE-RIS in subsequent coherence blocks. This leads to reduced training overhead, resulting in lower values of $\alpha$, which validates the efficiency of leveraging the estimates of the RIS-BS channel and the UE-RIS CCM for obtaining the phase-shifts. Furthermore, an additional overhead of signaling complexity is incurred to update the phase-shifts while using the I-CSI estimates that degrades the effective achievable rate. This makes optimizing the phase-shifts based on I-CSI estimates less appealing in practical scenarios.

\subsection{Effectiveness of Separate Channel Estimates}
We examine the performance of each estimate aside from the baselines of the capacity with phase-shifts derived from the exact channels and the phase-shifts derived from the exact RIS-BS channel and UE-RIS CCM. Fig. \ref{fig:perf_sep} shows the effectiveness of both neural networks \textit{Encoder} $\mathcal{F}$ and \textit{Encoder} $\mathcal{G}$ to estimate approximate posterior distributions to achieve desirable performance. We observe that at high SNR, both estimates, which are referred to as Perfect Channel - Estimated Covariance (\textbf{PC-ECov}) and Estimated Channel - Perfect Covariance (\textbf{EC-PCov}), can separately achieve the capacity with perfect RIS-BS channel and UE-RIS CCM. 
However, at low SNR, the covariance matrix estimates surpass the capacity achieved using channel estimates. This superiority is attributed to the highly sparse structure present in $\bm{d}$, originating from the clusters of AoAs and AoDs contributing to the UE-RIS channel $\bm{h}$. Moreover, the vector $\bm{d}$ has a lower dimension of $\bm{d}\in \mathbb{C}^N$ compared to the RIS-BS channel $\bm{G}\in\mathbb{C}^{M\times N}$, which facilitates the estimation of the non-sparse values.

\begin{table*}[!t]
    \centering
    \caption{Complexity analysis}
    \label{tab:complexity}
    \begin{tabular}{|c | c |c|}
        \hline
        \textbf{Model} & \textbf{FLOPs Encoder $\mathcal{F}$} & \textbf{FLOPs Encoder $\mathcal{G}$} \\
        \hline \hline
        JCE & $1087 M N_p + 3600N + 163740$ & $1087 MN_p + 3600 MN + 163740$ \\
        \hline
        JCCE & \makecell{$4MN_p N (MN_p+1) + 2MN_p +1087 (MN_p)^2$} & $4MN_p N (MN_p+1) + 2MN_p + 1087 (MN_p)^2$\\
        {} & $ + 600 N + 163740$ & $+ 3600MN + 163740$\\
        \hline
    \end{tabular}
\end{table*}

\subsection{Impact of Number of Coherence Blocks}
We assess the performance of the JCCE method in terms of the number of coherence blocks at $\text{SNR}=5\text{ db}$. Fig. \ref{fig:jcce_coh} depicts capacity and NMSE for different $N_b$ values. We note that increasing the number of coherence blocks is equivalent to increasing the number of realizations of the UE-RIS channel encompassed in the training signals from which we estimate the RIS-BS channel and the UE-RIS CCM. As shown in Fig. \ref{fig:cap_jcce_coh}, the JCCE does not require a large number of UE-RIS realizations to accurately estimate the covariance matrix which is of size $N \times N$. This efficiency is attributed to the VI framework embraced by JCCE, leveraging the low-rank structure of the UE-RIS CCM. Therefore, the JCCE significantly reduces the required number of UE-RIS channel realizations required to estimate the covariance matrix compared to the maximum likelihood estimator requires a number of estimates larger than $N$.
Furthermore, the NMSE of the RIS-BS channel consistently decreases with an increasing number of coherence blocks, as shown in \ref{fig:nmse_jcce_coh}. This improvement is attributed to the increase of training signal obtained during the $N_b$ coherence blocks, thereby resulting in a more precise estimation. Correspondingly, the NMSE of the vector $\bm{d}$ has a similar performance, depicting a reduction in error as the number of coherence blocks augments. This simulation assesses the effectiveness of the JCCE method and motivates the consideration of the UE-RIS CCM sparsity during estimation to reduce the training overhead.

\subsection{Complexity Analysis}

We provide the time-complexity analysis of the proposed methods. The neural networks are trained offline, therefore we evaluate only the inference mode, i.e., the forward propagation. The conventional method to evaluate the time-complexity of a neural network is the \textit{floating-point operations per second} (FLOPs) \cite{fredj2022distributed}. For any fully connected layer $L_i$ of input size $I_i$ and output size $O_i$ that follows a dropout layer of rate $1-r$ and a batch normalization layer, the number of FLOPs is given by
\begin{align}
    \text{FLOPs}(L_i) = 4r I + 2rI_i O_i.
\end{align}
Thus, the total number of FLOPs of the proposed neural network with 2 hidden layers yields
\begin{align}
    \text{FLOPs} &= \underbrace{4rI+2r I H_1}_{\text{input} \rightarrow L_1} + \underbrace{4 r H_1 + 2 r H_1 H_2}_{L_1 \rightarrow L_2} + \underbrace{4 H_2 + 2 H_2 O}_{L_2 \rightarrow \text{output}} \nonumber \\
    &= I\cdot(4r + 2rH_1) + O \cdot 2 H_2 \nonumber \\ 
    &\qquad + (2rH_1 + 2rH_1H_2 + 4H_2),
\end{align}
where $H_1$ and $H_2$ denote the size of the two hidden layers $L_1$ and $L_2$, respectively, $r$ is the dropout rate applied before the two hidden layers, $I$ represents the size of the input, and $O$ the size of the output. Note that the input of the encoders are real values, so the size of the input is multiplied by two considering the real and imaginary parts of the training signals. That is, for the JCE method, the input to the neural networks is of size $2MN_p$. Moreover, a preprocessing is performed to the training signal for the JCCE method, i.e., $\bm{\tilde{Y}}\bm{\tilde{Y}}^{\mathsf{H}}/N_b - \bm{I}_{MN_p}$, that adds a number of FLOPs equal to $4MN_pN(MN_p+1) + 2MN_p$. Table \ref{tab:complexity} compares the order of complexity of inference of the proposed VI-based methods. The computational complexity of the JCCE methods surpasses that of the JCE due to its handling of a larger number of observations, i.e., $M\times N_p \times N_b$. Fortunately, these computations primarily involve matrix multiplications, rendering efficient implementation feasible without incurring additional computational overhead, in contrast to model-based channel estimation methods necessitating optimization steps to estimate the channels.

\section{Conclusion}
Channel estimation poses a notable challenge for fully passive RIS-aided systems and the effectiveness of estimation schemes is dependent on the specific scenarios in which RIS systems are deployed. In this paper, we have tackled the CSI estimation problem in RIS-aided mmWave communication systems with fully-passive RIS elements using a VI-based framework to approximate the intractable posterior distribution of the channels with auxiliary distributions. In particular, we have proposed two different approaches addressing two scenarios in which the RIS is deployed. The first method, named JCE, separately estimates the UE-RIS and RIS-BS I-CSI that is suitable for scenarios with low mobility users. This method is useful for decoupling the cascaded channels and allows the identification of the channels' behavior in each part. However, its main limitation lies in its susceptibility to high training and signaling overhead as the UR-RIS channel becomes more dynamic for high mobile users. To overcome this challenge, leveraging the slow-varying nature of the RIS-BS I-CSI and the UE-RIS S-CSI, we have presented a second method, namely JCCE, that extends the VI-based framework used for JCE to estimate the RIS-BS channel and the UE-RIS CCM. Lastly, we have provided closed-form expressions of the phase-shifts given the obtained estimates for each use case considered in the methods.
We have showcased that sampling from the optimized auxiliary posterior distributions yields a capacity that is close to the one achieved with perfect CSI. Moreover, the JCCE provides an improvement of spectral efficiency through the reduction of the training overhead by relying on the slow-varying S-CSI of the UE-RIS channel rather than the I-CSI for the passive beamforming.
Several future directions can be adopted based on this work. For instance, a more physically consistent RIS modeling, where the elements of the RIS experience mutual coupling, which leads to a non-diagonal reflection matrix, can be studied. In addition, the multi-user scenario can be appropriately managed by employing identical phase-shifts for nearby users who share a similar covariance matrix.

\bibliographystyle{IEEEtran}
\bibliography{references}

\appendices
\appendix
In this section, we give a detailed derivation of the losses under the distributions investigated.

We derive the entropy of a complex Laplace random variable $z \sim \mathcal{CL}\bigl(m, b)$ with mean $m$ and scale $b$:
\begin{align}\label{eq:entropy_CL}
    H\bigl(q(z)\bigl) &= \int_{\mathbb{C}} - q(z) \log q(z)\; dz\nonumber\\
    &=\int_{\mathbb{C}} -\frac{1}{2\pi b^2} e^{-\frac{|z-m|}{b}} \log \frac{1}{2\pi b^2} e^{-\frac{|z-m|}{b}} dz\nonumber\\
    &=\log(2\pi b^2) + \int_{\mathbb{C}} \frac{|u|}{2\pi b^3} e^{-\frac{|u|}{b}} du \quad (u=z-m) \nonumber\\
    &=\log(2\pi b^2) + 2.
\end{align}
%\end{equation}

Next, we derive the closed-form of $\mathcal{L}^\mathsf{I-CSI}_3$ (Eq. (\ref{eq:loss_3_laplace})) with complex Laplace priors. In the first step, we compute the expectation over $\bm{h}^\mathsf{vir}$ where we denote $\bm{A}=\frac{\sqrt{\rho}}{MN^2} \bm{F}_M^{\mathsf{H}} \bm{G}^\mathsf{vir} \bm{F}_N^{\mathsf{H}} \text{diag}(\bm{v}_l) \bm{F}_N^{\mathsf{H}} x_l$ which is a constant with respect to $\bm{h}^\mathsf{vir}$:
\begin{align}
    \mathcal{L}_3^\mathsf{I-CSI} &= \sum_{l=1}^{N_p} \mathbb{E}_{\bm{h}^\mathsf{vir},\bm{G}^\mathsf{vir} \sim q_{\bm{\lambda}}(\bm{h}^\mathsf{vir},\bm{G}^\mathsf{vir}|\bm{Y})} \Bigr[(\bm{y}_l - \bm{A}\bm{h}^\mathsf{vir})^{\mathsf{H}} \nonumber\\
    &\qquad\times(\bm{y}_l - \bm{A}\bm{h}^\mathsf{vir}) \Bigr] + C_1\nonumber\\
    &= \sum_{l=1}^{N_p} \mathbb{E}_{\bm{G}^\mathsf{vir} \sim q_{\bm{\lambda}_2}(\bm{G}^\mathsf{vir}|\bm{Y})} \Bigr[ \trace \bigl( \bm{A} \bm{\Lambda} \bm{A}^{\mathsf{H}}\bigl)\nonumber\\
    &\qquad+ (\bm{y}_l - \bm{A} \bm{m})^{\mathsf{H}} (\bm{y}_l - \bm{A} \bm{m}) \Bigr] + C_1,
\end{align}
where $C_1$ is a constant, $\bm{m}$ a vector of means of $\bm{h}^\mathsf{vir}$ following $q_{\bm{\lambda}_1}(\bm{h}^\mathsf{vir}|\bm{Y})$ distribution and $\bm{\Lambda} = \mathbb{E}_{\bm{h}^\mathsf{vir} \sim q_{\bm{\lambda}_1}(\bm{h}^\mathsf{vir}|\bm{Y})}[(\bm{h}^\mathsf{vir}-\bm{m})(\bm{h}^\mathsf{vir}-\bm{m})^{\mathsf{H}}]$ is the covariance matrix of $\bm{h}^\mathsf{vir}$. The latter is a diagonal matrix with a main diagonal containing the variances of the elements. The variance of a complex Laplace is defined as follows:
\begin{align}
    \text{Var}(z) &=\int_\mathbb{C} \frac{|z-m|^2}{2\pi b^2} e^{-\frac{|z-m|}{b}} dz\nonumber\\
    &=\int_\mathbb{C} \frac{|u|^2}{2\pi b^2} e^{-\frac{|u|}{b}} du \quad (\text{Substitution } u=z-m)\nonumber\\
    &=\int_0^{2\pi} \int_0^\infty \frac{r^2}{2\pi b^2} e^{-\frac{r}{b}} r\;dr\;d\theta \quad (\text{polar coordinates})\nonumber\\
    &=6b^2.
\end{align}
Hence, the covariance matrix $\bm{\Lambda}$ is expressed as follows:
\begin{equation}
    \bm{\Lambda}_{i,j} = 6 \;\text{diag}(\bm{b})^2.
\end{equation}
To compute $\bm{G}^\mathsf{vir}$, we define a constant matrix $\bm{C}= \frac{\sqrt{\rho}}{MN^2} \bm{F}_N^{\mathsf{H}} \text{diag}(\bm{v}_l) \bm{F}_N^{\mathsf{H}} x_l$, i.e, $\bm{A}=\bm{F}_M^{\mathsf{H}} \bm{G}^\mathsf{vir} \bm{C}$. Hence, we get:
\begin{align}
    \mathcal{L}^\mathsf{I-CSI}_3 &=  \sum_{l=1}^{N_p} \mathbb{E}_{\bm{G}^\mathsf{vir} \sim q(\bm{G}^\mathsf{vir}|\bm{Y})} \Bigr[ \trace \bigl( \bm{A}^{\mathsf{H}} \bm{A} \bm{\Lambda}\bigl) \nonumber\\
    &\quad+ (\bm{y}_l - \bm{A} \bm{m})^{\mathsf{H}} (\bm{y}_l - \bm{A} \bm{m}) \Bigr]+C_1\nonumber\\
    &= \sum_{l=1}^{N_p} \mathbb{E}_{\bm{G}^\mathsf{vir}\sim q(\bm{G}^\mathsf{vir}|\bm{Y})} \Bigr[ M \trace\bigl(\bm{C}^{\mathsf{H}} {\bm{G}^\mathsf{vir}}^{\mathsf{H}} \bm{G}^\mathsf{vir} \bm{C} \bm{\Lambda} \bigl)  \nonumber\\
    &\quad+ (\bm{y}_l - \bm{F}_M^{\mathsf{H}} \bm{G}^\mathsf{vir} \bm{C} \bm{m})^{\mathsf{H}} (\bm{y}_l - \bm{F}_M^{\mathsf{H}} \bm{G}^\mathsf{vir} \bm{C} \bm{m})\Bigr] + C_1.
\end{align}
Then we use the property $\mathbb{E}_{\bm{G}^\mathsf{vir}}[{\bm{G}^\mathsf{vir}}^{\mathsf{H}} \bm{G}^\mathsf{vir}] = \bm{Q} + \bm{M}^{\mathsf{H}} \bm{M}$ where $\bm{Q}=\mathbb{E}_{\bm{G}^\mathsf{vir}}[(\bm{G}^\mathsf{vir}-\bm{M})^{\mathsf{H}}(\bm{G}^\mathsf{vir} -\bm{M})]$ is the covariance matrix over the columns of $\bm{G}^\mathsf{vir}$. $\bm{Q}$ is a diagonal matrix since the elements $\bm{G}^\mathsf{vir}_{i,j}$ are assumed to be independent which makes the columns independent as well and the elements on the diagonal are given by:
\begin{equation}
\begin{split}
    \bm{Q}_{i,i} &= \sum_{m=1}^M \text{Var}(\bm{G}^\mathsf{vir}_{m,i})=\sum_{m=1}^M 6 \bm{B}_{m,i}^2.
\end{split}
\end{equation}
Therefore, we have:
\begin{equation}
\begin{split}
    \mathcal{L}_3^\mathsf{I-CSI} &=  \sum_{l=1}^{N_p} \Bigr[ M \trace\bigl(\bm{C}^{\mathsf{H}} \bm{Q} \bm{C} \bm{\Lambda} \bigl) + M \trace\bigl(\bm{C}^{\mathsf{H}} \bm{M}^{\mathsf{H}} \bm{M} \bm{C} \bm{\Lambda} \bigl)\\
    &\qquad+(\bm{y}_l - \bm{F}_M^{\mathsf{H}} \bm{M} \bm{C} \bm{m})^{\mathsf{H}} (\bm{y}_l - \bm{F}_M^{\mathsf{H}} \bm{M} \bm{C} \bm{m}) \nonumber\\
    &\qquad+ M  \bm{m}^{\mathsf{H}} \bm{C}^{\mathsf{H}} \bm{Q} \bm{C} \bm{m}\Bigr] + C_1.
\end{split}
\end{equation}

\end{document}